\documentclass[a4paper,11pt]{article}
\pdfoutput=1 

\usepackage{jheppub} 

\usepackage[T1]{fontenc} 
\usepackage{array}
\usepackage{graphicx,rotate,multicol,caption}
\newcommand{\bea}{\begin{eqnarray}}
\newcommand{\beq}{\begin{equation}}
\newcommand{\eea}{\end{eqnarray}}
\newcommand{\eeq}{\end{equation}}

\newcommand{\vep}{\varepsilon}

\newcommand{\nn}{\nonumber}

\newcommand{\lsim}{\raise0.3ex\hbox{$\;<$\kern-0.75em\raise-1.1ex\hbox{$\sim\;$}}}
\newcommand{\gsim}{\raise0.3ex\hbox{$\;>$\kern-0.75em\raise-1.1ex\hbox{$\sim\;$}}}
\newcommand{\eq}[1]{Eq.~(\ref{#1})}
\newcommand{\unity}{{\hbox{1\kern-.8mm l}}}

\title{\boldmath Slepton Non-Universality in the Flavor-Effective MSSM}

\author[a]{M. Luisa L\'opez-Ib\'a\~nez,}
\author[a]{Aurora Melis,}
\author[a,b]{M. Jay P\'erez}
\author[a]{and Oscar Vives}

\affiliation[a]{Departament de F\'{i}sica T\`{e}orica, Universitat de Val\`{e}ncia and IFIC, Universitat de Val\`{e}ncia-CSIC  \\
	Dr. Moliner 50, E-46100 Burjassot (Val\`{e}ncia), Spain}
\affiliation[b]{Valencia State College, Osceola Campus \\
	1800 Denn John Ln, Kissimmee, FL, USA}

\emailAdd{m.luisa.lopez-ibanez@uv.es}
\emailAdd{aurora.melis@ific.uv.es}
\emailAdd{mperez75@valenciacollege.edu}
\emailAdd{oscar.vives@uv.es}

\preprint{FTUV-17-1005.3978, IFIC-17-46}

\abstract{Supersymmetric theories supplemented by an underlying flavor-symmetry $\mathcal{G}_f$ provide a rich playground for model building aimed at explaining the flavor structure of the Standard Model. In the case where supersymmetry breaking is mediated by gravity, the soft-breaking Lagrangian typically exhibits large tree-level flavor violating effects, even if it stems from an ultraviolet flavor-conserving origin. Building on previous work, we continue our phenomenological analysis of these models with a particular emphasis on leptonic flavor observables. We consider three representative models which aim to explain the flavor structure of the lepton sector, with symmetry groups $\mathcal{G}_f = \Delta(27)$, $A_4,$ and $S_3$.}

\begin{document} 
\maketitle
\flushbottom
\section{Introduction}
\label{s:intoduction}

As the LHC marches onward in its search for hints of physics beyond the Standard Model, the community eagerly waits. Unfortunately, despite the overwhelming evidence for its need in order to explain open questions, such as the nature of dark matter, the stability of the Higgs mass with respect to higher scales, the origin of the Baryon asymmetry in the universe, amongst others, New Physics (NP) continues to elude us. We should not  despair however, as the LHC, along with a robust set of other dedicated experiments, will continue to probe new corners of parameter space where NP could be hiding. At the same time, as our ``first-guess'' models come increasingly under pressure, it is worth pausing to consider alternative methods or observables which may help to further constrain them and extend the reach of the LHC.  

A class of such models, popular for their ability to shed light on several of the open questions in particle physics, are supersymmetric extensions of the Standard Model. Its simplest incarnation, the Minimal Supersymmetric Standard Model (MSSM), has many virtues: a possible dark matter candidate; new sources of CP violation; a mechanism for stabilizing the mass of the Higgs; the possibility for unification of the fundamental forces. However the non-observance by the LHC in Runs 1 and 2 of any of its predicted superpartners is beginning to constrain such a minimal realization of supersymmetry, pointing to a mass scale of the new predicted particles which may be heavier than naively expected. In the scenario where supersymmetry is indeed realized by nature, but out of reach of current colliders, we should look for further ways to probe or constrain the large parameter space available in the MSSM.   

One curious legacy of the Standard Model (SM) is its rich flavor structure, which has historically \cite{Glashow:1970gm} proven invaluable for and complementary to direct searches for sniffing out new particles. Yet, understanding the peculiar mass and mixing pattern of the fundamental fermions remains one of the biggest puzzles of the SM. Despite a wealth of ideas and models put forth by the theory community, a convincing solution to this puzzle is still missing. Among the proposed ideas, the use of flavor symmetries, both continuous and discrete, remains a popular tool for model builders. This avenue has been especially explored in the lepton sector, where the suggestive form of the Pontecorvo-Maki-Nakagawa-Sakata (PMNS) matrix has led to several ans{\"a}tze for its decomposition in terms of primitive ``bare" mixing matrices, which give leptonic mixing angles close to their measured values. In most models, the aim is to motivate these special angles through the Clebsch-Gordan (CG) coefficients of a symmetry group, and moreover, to predict the as yet unmeasured parameters of the leptonic sector: the Dirac CP violating phase, the quadrant of the atmospheric angle, and the neutrino mass ordering. 

Unfortunately, a definitive picture has failed to emerge from the large number of present models (for recent reviews, see\cite{Babu:2009fd,Altarelli:2010gt}). One well-known problem at the level of the SM is that we cannot fully reconstruct the fundamental flavor parameters of the SM Lagrangian, the Yukawa matrices. In this regard, NP models which predict new flavor interactions in addition to new particles are particularly interesting, as they are bound to shed additional light (right-handed mixings, etc.) on the flavor puzzle regardless of their original motivations.  
  
The MSSM contains a wealth of such new flavor interactions in its soft-breaking sector. Although, in all generality the MSSM contains a host of unknown parameters in the flavor sector, in a previous work \cite{Das:2016czs} we explored a specific class of \emph{predictive} models where the MSSM emerges as an effective theory from an ultraviolet flavor-symmetric theory. These models :

\begin{itemize}

\item Arise from a superpotential which is invariant under a given flavor symmetry $\mathcal{G}_f$, spontaneously broken at a scale $\Lambda_f$. After the breaking of $\mathcal{G}_f$, new effective operators, \`a la Froggatt-Nielsen (FN), contribute to the low-energy superpotential. Similar effective operators contribute to the soft-breaking Lagrangian. 

\item Mediation of Supersymmetry breaking to the visible sector is assumed to occur through interactions at a scale $\Lambda_S  \gg \Lambda_f$, so that the soft-breaking terms, and, more exactly, the visible sector operators giving rise to the soft-breaking terms, respect $\mathcal{G}_f$. An illustrative example of such a mediation scheme, which we will assume for simplicity, is gravity mediation.
\end{itemize} 

Under these conditions, these models contain \emph{tree-level} flavor violating effects, arising from the mismatch between the order one coefficients of their supersymmetric and corresponding supersymmetry-breaking supergraphs after integrating out the mediator fields at $\Lambda_f$. In addition, as the flavor parameters \footnote{With the exception of the usual unknown order-one parameters.} are fixed by the structure of the superpotential, these models are minimal, depending only on the traditional supergravity input parameters $m_0$, $m_{1/2}$, $A_0$, $\tan \beta$, and $\mu$. This minimality and calculabity of these models makes them interesting in their own right, and especially amenable to constraints from flavor observables; in many cases extending beyond the reach of direct searches at the LHC.

In this work, we continue our investigation of this class of models \cite{Das:2016czs,Calibbi:2012yj,Antusch:2011sq}, with a particular emphasis on constraints coming from leptonic flavor observables such as $\mu \rightarrow e \gamma$, $\mu \rightarrow eee$, and $\mu - e$ conversion, although for completeness we scan each model over all relevant flavor observables to obtain the strongest constraints. We look at three representative models available in the literature, based on the symmetry groups $\Delta(27)$, $A_4$ and $S_3$. 

Our paper is organized as follows. We begin in Sec.~\ref{overview} with a short review of the mechanism presented in  \cite{Das:2016czs}, giving generic formulas applicable for any of the class of models under investigation. In Secs.~\ref{del27model}-\ref{s3model} we apply these general formulae to specific models found in the literature based on the flavor groups $\Delta(27)$, $A_4$, and $S_3$. These sections are self-contained, including the relevant phenomenological analyses and results for each group. Finally, we conclude in Sec.~\ref{conclusions} with brief remarks on our general results and future outlooks for extensions of this work. 


\section{A Review of the Mechanism}
\label{overview}

In this section we review and update the results of our previous work  \cite{Das:2016czs}, demonstrating that in SUSY models augmented with a flavor symmetry spontaneously broken at a scale $\Lambda_f \leq \Lambda_S$, flavor violation in the soft-breaking terms is generically present in the low-energy effective theory. This remains true even starting with completely flavor blind soft-breaking in the full theory and runs contrary to the naive expectation that the soft terms, being controlled by the flavor symmetry, should be diagonalized by the same rotations which diagonalize the Yukawa couplings.
This mismatch between the Yukawa or Kinetic mixing matrices and their corresponding soft-term structures stems from the different ways in which SUSY breaking may be inserted in the full theory diagrams, giving rise to a single coupling in the low-energy effective theory. 

Supersymmetry breaking can be represented by the insertion of a chiral background superfield, a spurion $X$, which is assumed to obtain a vacuum value largely along its supersymmetry breaking component $\langle F_X \rangle \gg \langle X \rangle$. Although not necessary, we will make the simplifying assumption in this work that this spurion is the only source of SUSY breaking and couples universally to the visible sector. 

\begin{figure} 
\center
\includegraphics[width=0.5\textwidth]{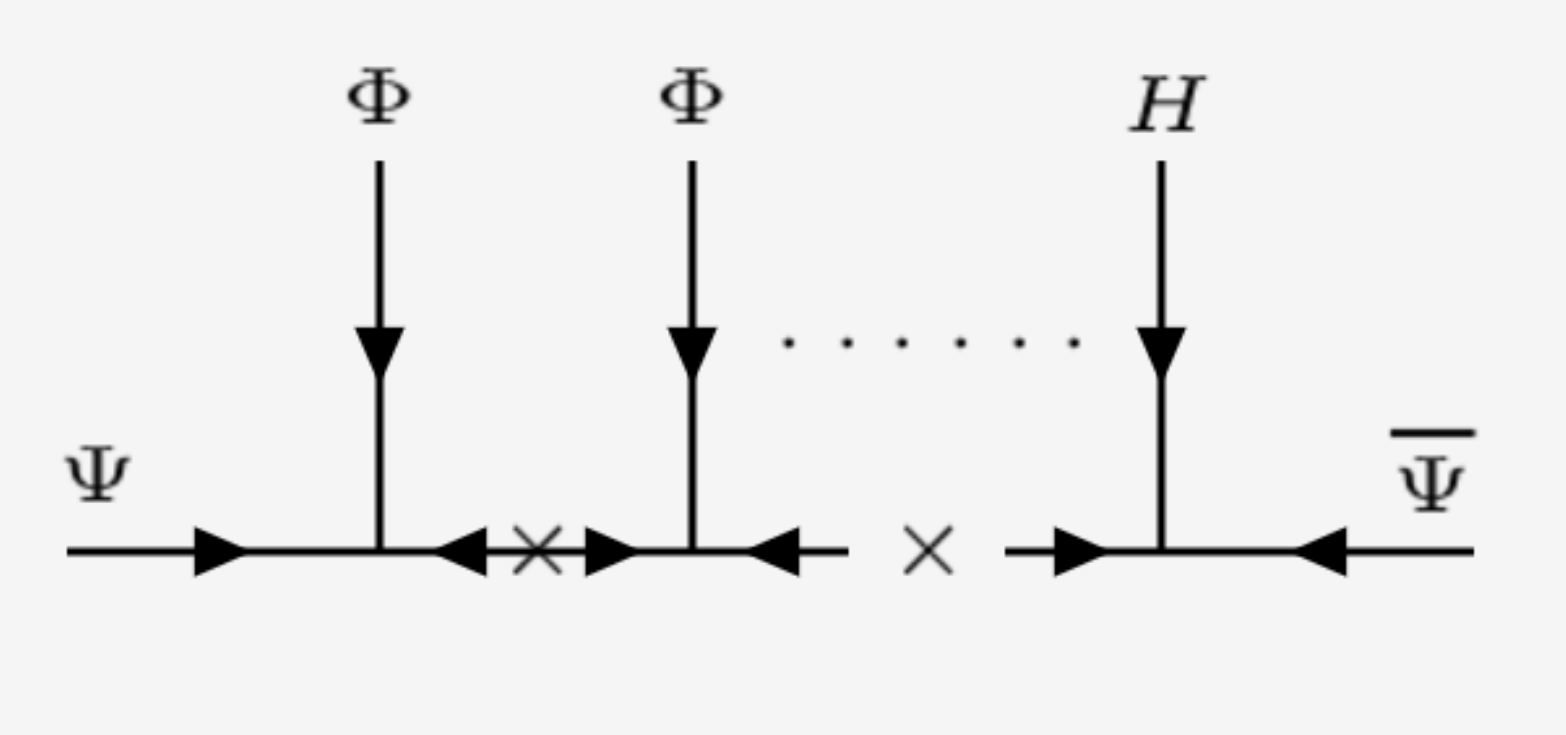} \\
\includegraphics[width=0.35\textwidth]{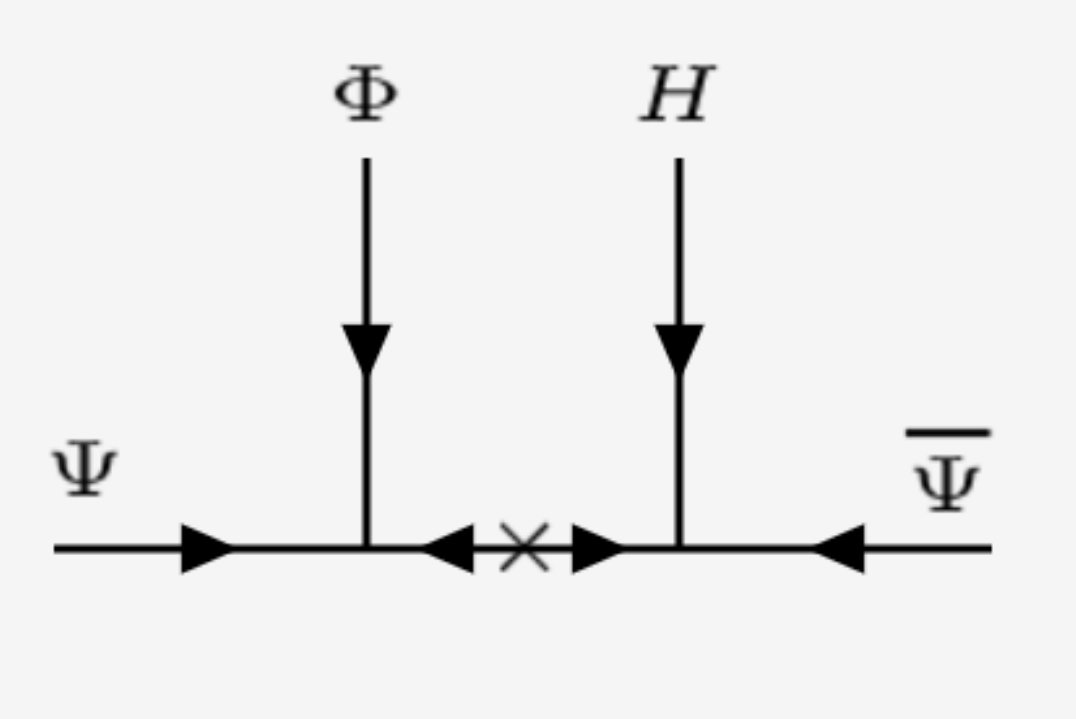}  %
\caption{A supergraph depiction of the corrections to the superpotential represented by Eq.~\ref{eq:suppot}. An example for $n=2$ involving a single Flavon insertion is given below. The internal lines are heavy messengers, and the cross denotes a supersymmetric mass insertion $M$.}
\label{fig:suppot}
\end{figure}

This mismatch between the soft-breaking terms and the superpotential or K\"ahler potential is manifest in terms of the FN fields in the full theory. Corrections to the low-energy superpotential $W$ and K\"ahler potential $K$ are generated below the flavor scale, $\Lambda_f$. These corrections stem from non-renormalizable operators containing an appropriate number of flavon insertions, generated by integrating over the appropriate heavy messengers in the underlying theory, which, in the case of the superpotential, may write schematically as, 
\begin{equation} \label{eq:suppot}
W = W_{\rm ren} + \Psi ~ \overline \Psi~ H \left( \sum_{n=1}^{\infty} x_n \left(\frac{\langle \Phi \rangle }{M} \right)^n \right), 
\end{equation}
where $\Psi$ ($\overline \Psi$) denotes any of the left-handed (right-handed conjugate) MSSM superfields, $H$ denotes the SM Higgs field, $\langle \Phi \rangle $ the vacuum expectation values (VEVs) of any of the flavons or heavy Higgses, $M$ the heavy mass scale $\sim \Lambda_f$ of the messengers and $x_n$ is a numerical coefficient depending on the charges of the fields. These corrections may be represented schematically in terms of the supergraphs which generate them, as shown in Fig.~\ref{fig:suppot}. 

In addition to correcting to the superpotential, similar supergraphs will generate the so called $A$-terms in the soft Lagrangian upon inserting a soft-breaking term at any internal point in the diagram, which can be represented by the insertion of a spurion field $X$ with non-vanishing $F$-term, $F_X$. Assuming a universal SUSY breaking, these universal corrections in the full theory are of the form
\begin{equation} \nonumber
{\cal L}_{\rm soft} \sim \frac{F_X}{M_{\rm Pl}} \times W_{\rm ren} \equiv m_0 \times W_{\rm ren}
\end{equation} 
In terms of our supergraph language, this corresponds to attaching an external line involving the spurion $X$ to each of the vertices in a given supergraph. 

From here, it is evident that, after integrating out the heavy fields in the Lagrangian to obtain the low-energy effective theory, the different ways to couple the spurion field produce a mismatch between the $A$ terms and their corresponding Yukawa matrices. For a given supergraph which generates an entry in the Yukawa matrix, we have multiple ways to generate the corresponding $A$ term, one for each insertion of the spurion $X$ at a given vertex. This mismatch may be easily written in terms of the operator dimension which generates the given entry in the Yukawa matrices of the superpotential. Given an operator with $N$ $\Phi$ insertions, we have $2N+1$ possible $X$ insertions; 2 for each $\Phi$ and mass-insertion vertex, plus one additional for the vertex involving the Higgs. Generically, this implies that for a Yukawa entry $Y_{ij}$ generated by $N$ Flavon insertions, 

\begin{equation} \label{eq:amismatch}
A_{ij} \sim (2N+1)~ a_0\; Y_{ij}
\end{equation} 

where $a_0=k\, m_0$. 
As in FN models each entry in the Yukawa matrix is generated at a different order, the individual entries in the $A$ matrices will contain different order one coefficients, and not be directly proportional to the Yukawa matrices. Performing a rotation of the superfields and going to the Super-CKM basis, the $A$ terms will not be diagonalized, their off-diagonal terms contributing at tree-level to flavor violating observables.

Similar considerations hold for the K\"ahler potential. Below $\Lambda_f$, corrections to the K\"ahler potential are generated when integrating over the heavy messengers. In the case of a single flavon, as in the case Abelian models, it can be written schematically as, 
\begin{equation} \label{eq:kahler}
(K_\Psi)_{ij} = \Psi_i ~\Psi^{\dagger}_j \left(\delta_{ij} + \sum_{n,m} c_{ij}^{(n,m)} \left(\frac{\Phi}{M} \right)^n  \left(\frac{\Phi^\dagger}{M} \right)^m \right),
\end{equation}
where, for the leading terms, $c_{ij}^{(n,m)} = \delta_{m,0}~ \delta (q_i + q_j - n)$ if $(q_i+q_j) > 0$ and  $c_{ij}^{(n,m)} =\delta_{n,0}~ \delta (q_i + q_j - m)$ if $(q_i+q_j) < 0$.

In the case of several flavon fields in complex representations of $\mathcal{G}_f$, as is the case of typical non-Abelian models, the {\bf leading} contributions appear in the form $\Phi_r \Phi_r^{\dagger}$\footnote{Depending on the model, there may exist other contributions, including even non-hermitian combinations of fields, if they are neutral under the different charges. However, they are usually sub-leading with respect to $\Phi_r \Phi_r^{\dagger}$.},
\begin{equation} \label{eq:kahler2}
(K_\Psi)_{ij} = \Psi_i ~\Psi^{\dagger}_j \left(\delta_{ij} + \sum_{r,n} c_{ij}^{r,n} \left(\frac{\Phi_r \Phi_r^\dagger}{M^2} \right)^n + \dots \right),
\end{equation}

\begin{figure} 
\center
\includegraphics[width=0.5\textwidth]{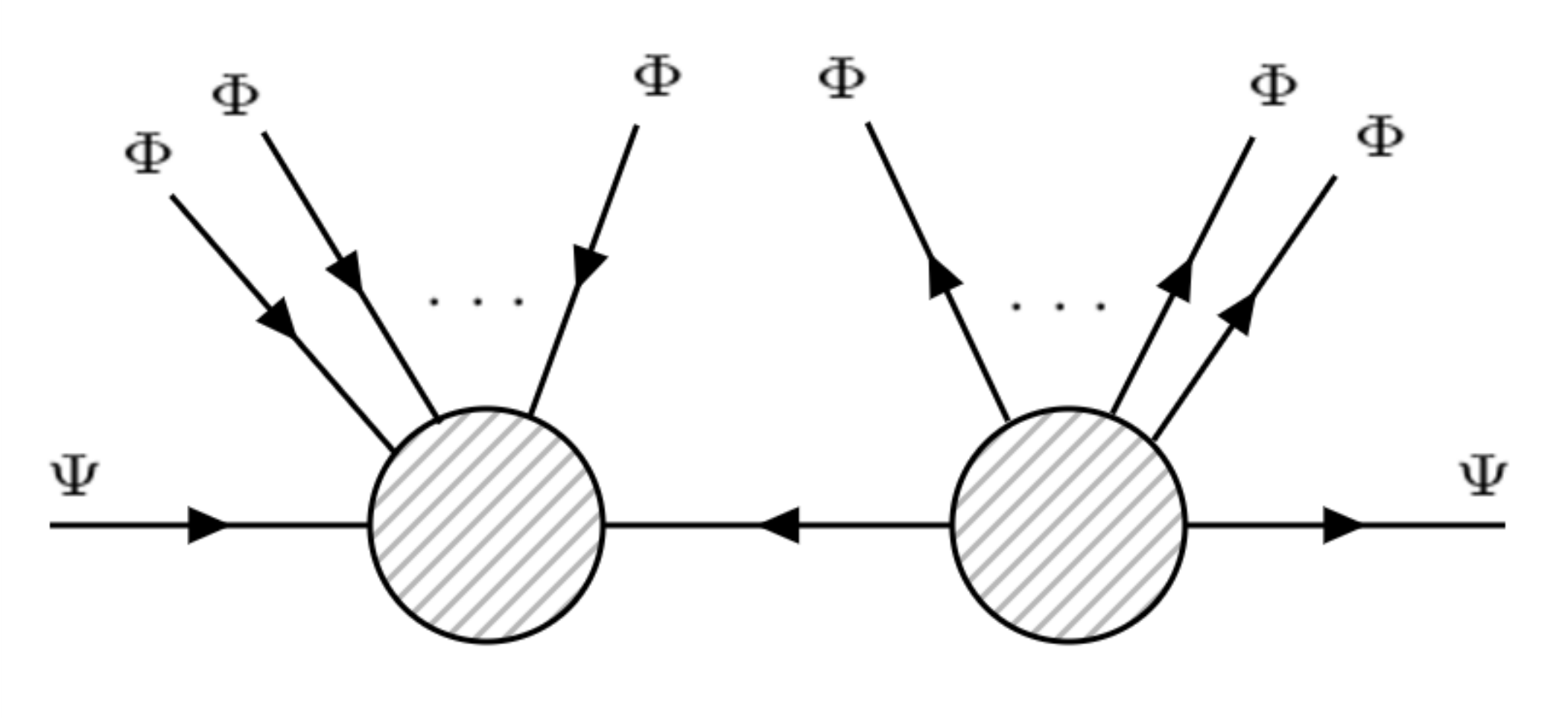} \\
\includegraphics[width=0.6\textwidth]{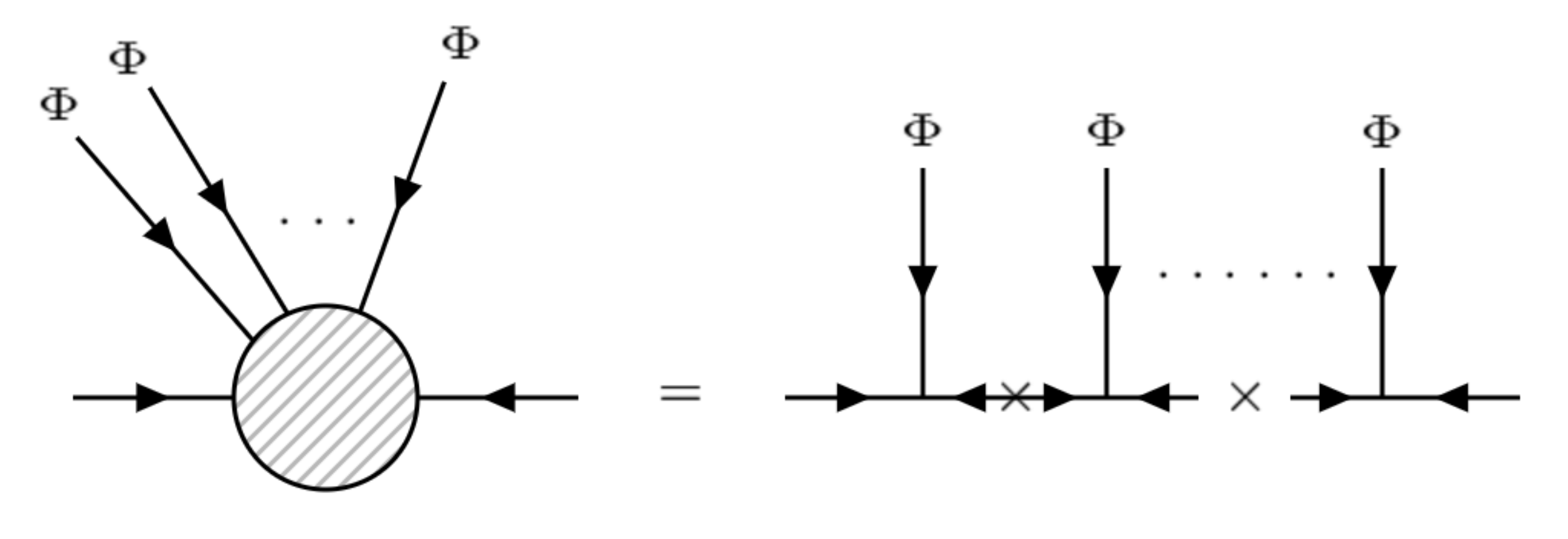}  %
\caption{Supergraphs which correct the K\"ahler potential.}
\label{fig:kahler}
\end{figure}

Again, this can be depicted in terms of supergraphs, where now superfields may both enter (undaggered) or leave (daggered) a given vertex. The leading corrections, those that do not contain derivatives or additional suppressions of $M$, are all of the form shown in Fig.~\ref{fig:kahler}, with one internal line a superpropagator of a given messenger connecting ``bubbles" of $\Phi$'s involving only mass insertions in the internal lines. We may therefore organize the corrections generated by a given supergraph by the number of incoming ($N_{\rm in}$) and outgoing ($N_{\rm out}$) $\Phi$'s.

\begin{figure} 
\center
\includegraphics[width=0.4\textwidth]{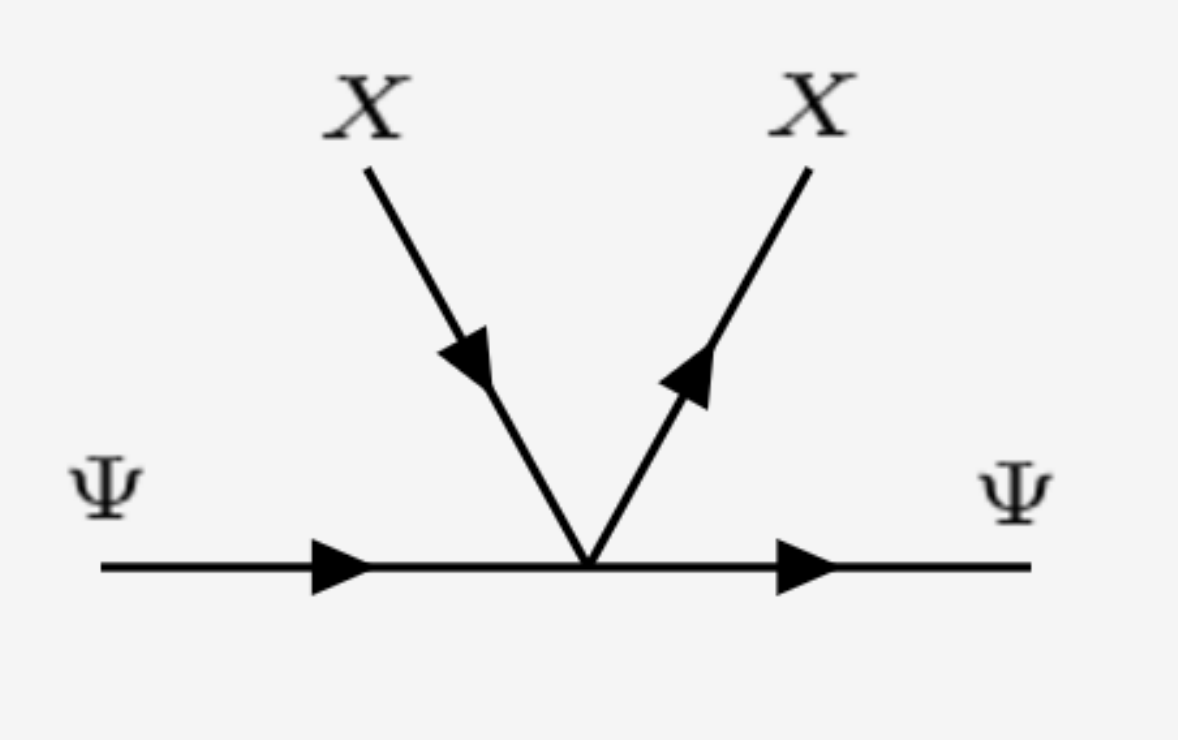}        
\caption{Diagonal contribution to the soft masses of a given superfield $\Psi$.}
\label{fig:softmassdiag}
\end{figure}

\begin{figure} 
\center
\includegraphics[width=0.5\textwidth]{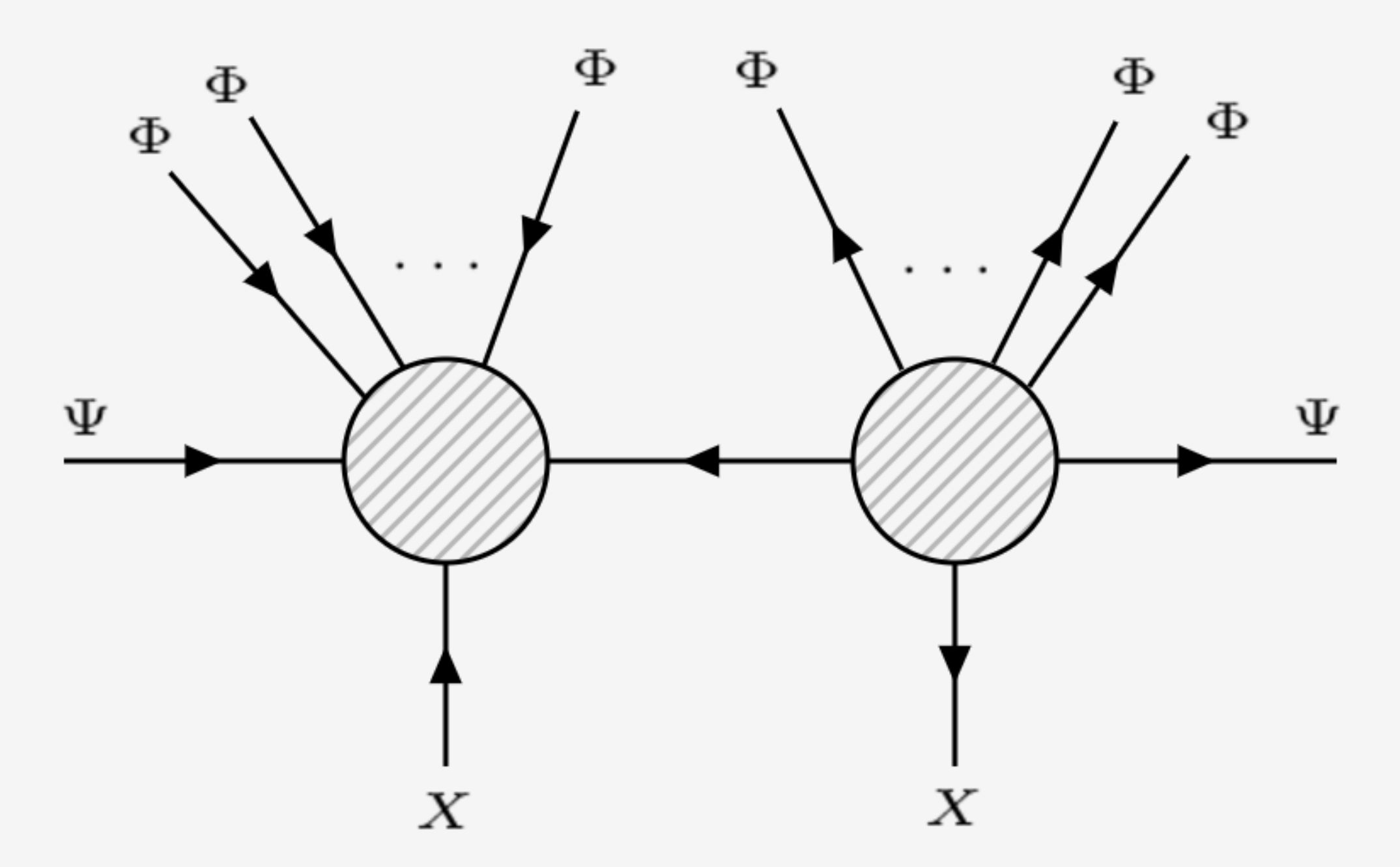}        
\caption{Schematic contribution to the soft mass of a superfield $\Psi$.}
\label{fig:softmasses}
\end{figure}

A given supergraph of this form will generate soft masses for the corresponding scalars $\tilde \Phi$ when coupled to the supersymmetry breaking combination $\langle F_X \rangle \langle F_X \rangle^{\dagger}$, as shown for the diagonal contribution in Fig.~\ref{fig:softmassdiag}. For a supergraph of the form of Fig.~\ref{fig:kahler}, we have have two ways to attach the spurion combination $X X^{\dagger}$, either as in Fig.~\ref{fig:softmassdiag} to an internal superpropagator, or with $X$ attached to one of the incoming $\Phi$ vertices and $X^{\dagger}$ attached to one of the outgoing $\Phi$ vertices, as shown in Fig.~\ref{fig:softmasses}.


\begin{figure} 
\center
\includegraphics[width=0.7\textwidth]{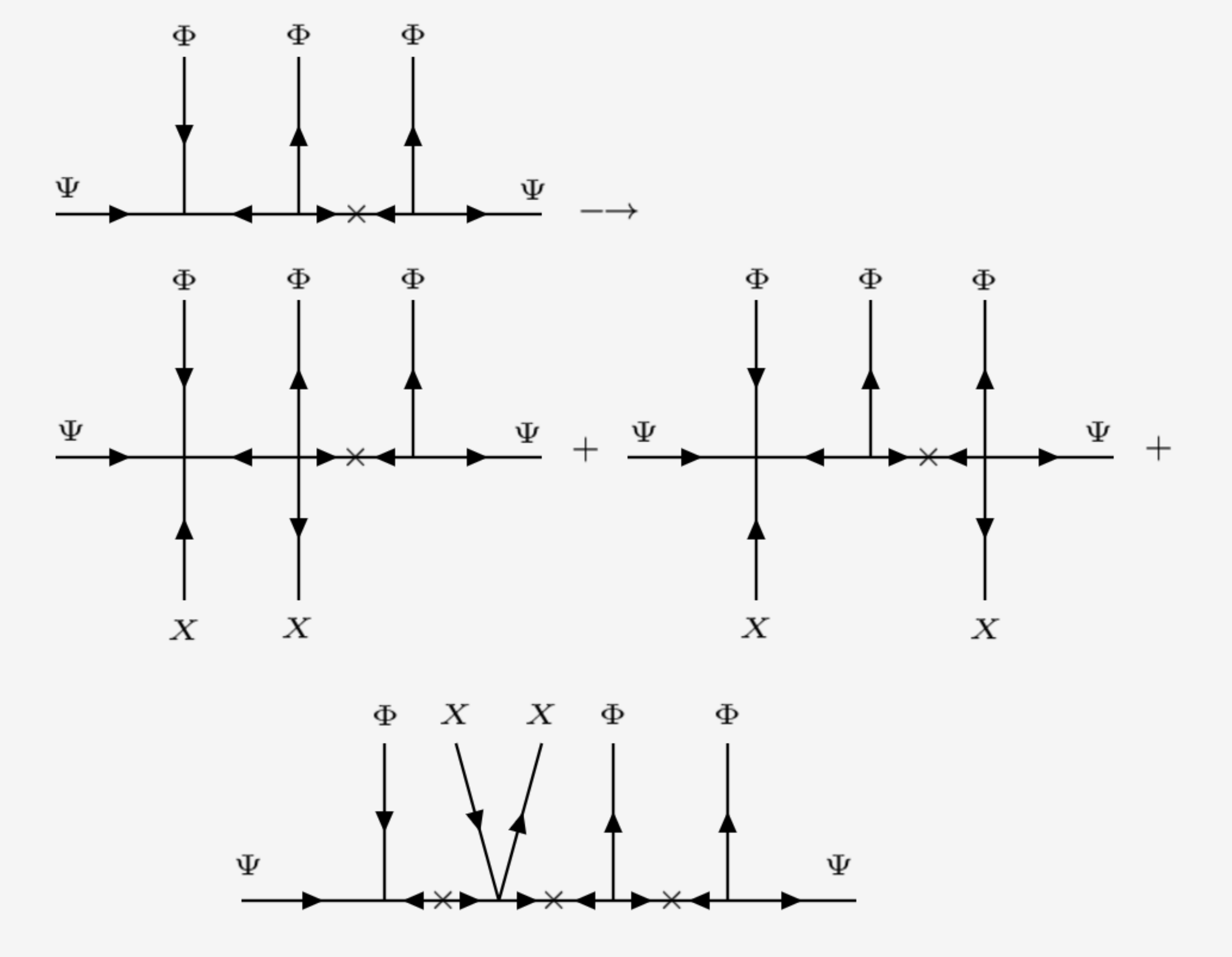}        
\caption{An example of the mismatch factor in the soft masses for $N_{\rm in}=1$, $N=3$.}
\label{fig:kahlercorex1}
\end{figure}

As there are $N_{\rm in}$ ways to attach $X$ to a given incoming vertex and $N_{\rm out}$ ways to attach $X^{\dagger}$ to an outgoing vertex, plus an additional graph involving the correction to the internal superpropagator, we find that the mismatch factor between the soft mass matrices and the K\"ahler matrices can be written in terms of the total number of Flavon insertions $N= N_{\rm in}+N_{\rm out}$ and the number of incoming Flavon insertions $N_{\rm in}$,

\begin{equation} \label{eq:kmismatch}
(m^2_{\Psi})_{ij} \sim f\; m_0^2 \cdot (K_\Psi)_{ij}, \qquad f ~=~ N_{\rm in} \cdot N_{\rm out}+1 ~=~ N_{\rm in} \cdot (N - N_{\rm in}) +1 .
\end{equation}
As a concrete example, we show the case with $N_{\rm in}=1$, $N=3$ in Fig.~\ref{fig:kahlercorex1}, for which Eq.~\ref{eq:kmismatch} gives $f=3$. 

Eqs.~\ref{eq:amismatch} and \ref{eq:kmismatch} are useful in the sense that without knowing precisely the underlying theory, the mismatch factors can be quickly calculated solely in terms of the number of Flavon insertions, or alternatively, the operator dimension at which a given Yukawa entry is generated. Once these mismatch factors are known and the soft-matrices given, rotations of the superfields, first to canonically normalize \cite{King:2004tx} and then to diagonalize the Yukawa matrices, may be performed. 

It is worth noting that even if the leading non-universal contributions in the soft-mass matrix are proportional to the K\"ahler matrix, flavor changing entries are generically present in the SCKM basis. In this case, the diagonalization of the K\"ahler matrix also diagonalizes the soft-mass matrix, but the rescaling of the diagonal K\"ahler elements does not  
eliminate the diagonal elements in the soft-mass matrices if $f\neq 1$; off-diagonal elements will always reappear when going to the SCKM basis. 

As an illustrative example, consider a simplified non-Abelian model with two flavons. The non-universal corrections to the K\"ahler potential and soft-mass matrices would be proportional,
\begin{equation} \label{eq:km2ex}
 K_{ij} = \delta_{ij} + c_1 \left(\frac{\Phi_1 \Phi_1^\dagger}{M^2}\right) +  c_2 \left(\frac{\Phi_2 \Phi_2^\dagger}{M^2}\right), \qquad   (m_{ij})^2 = m_0^2 \left(\delta_{ij} + 3 c_1 \left(\frac{\Phi_1 \Phi_1^\dagger}{M^2}\right) +  3 c_2 \left(\frac{\Phi_2 \Phi_2^\dagger}{M^2}\right)\right) .
\end{equation}
Taking $\Phi_1=(0,1)$ and $\Phi_2=(\vep,\vep)$, it is clear that both matrices are diagonalized with the same rotation $U$, but the rescaling of the K\"ahler, $N^{1/2}$, does not reabsorb the non-universal diagonal elements in the soft mass matrix,
\begin{equation} \label{eq:km3}
N^{1/2} U^\dagger K_{ij} U N^{1/2} = \unity
\hspace{2.cm}
 N^{1/2} U^\dagger (m_{ij})^2 U  N^{1/2} \simeq m_0^2 \left(\begin{array}{cc}
               1 + 2~a_2\,\vep    &  0 \\
               0 & 1 + b_1 + b_2\vep \,   
               \end{array} \right),
\end{equation}
with $b_1 \simeq 2 c_1/(1+ c_1)$ and $b_2 \simeq 2 c_2/(1 + c_1^2)$. Thus, as stated before, when diagonalizing the Yukawa matrix to go to the SCKM basis, the new rotation $V \sim \cal{O} (\vep)$, will introduce again off-diagonal terms in the soft-mass matrices.

These off-diagonal entries of the $A$ terms and soft masses are very relevant in performing phenomenological analyses of given models. By subjecting them to the appropriate flavor constraints, like those collected in Table \ref{tab:LFVbounds} for leptonic processes, complementary bounds to high-energy colliders can be set. 

{\renewcommand{\arraystretch}{1.5}
\begin{table}[t!]
\centering 
\begin{tabular}{|c|c|c|}
\hline
LFV process & Current Bounds \cite{Olive:2016xmw}    & Future Bounds     \\
\hline 
BR($\mu  \to e \gamma$)     & $4.2 \times 10^{-13}$  & $4 \times 10^{-14}$ \cite{Baldini:2013ke} \\
BR($\tau \to e \gamma$)     & $3.3 \times 10^{-8}$   & \qquad $10^{-9}$ \cite{Aushev:2010bq} \\
BR($\tau \to \mu \gamma$)   & $4.4 \times 10^{-8}$   & \qquad $10^{-9}$ \cite{Aushev:2010bq} \\
BR($\mu  \to e e e$)        & $1.0 \times 10^{-12}$  & ~~\quad $10^{-16}$ \cite{Blondel:2013ia} \\
BR($\tau \to e e e$)        & $2.7 \times 10^{-8}$   & \qquad $10^{-9}$ \cite{Aushev:2010bq} \\
BR($\tau \to \mu \mu \mu$)  & $2.1 \times 10^{-8}$   & \qquad $10^{-9}$ \cite{Aushev:2010bq} \\
BR($Z \to e \mu$)           & $7.5 \times 10^{-7}$   & - \\
BR($Z \to e \tau$)          & $9.8 \times 10^{-6}$   & - \\
\hline
\end{tabular}
\caption{\label{tab:LFVbounds}Relevant Lepton Flavour Violating (LFV) processes considered in our analysis.}
\end{table}}

Finally, an additional consideration comes from the stability of the vacuum. As shown in  \cite{Casas:1996de}, the requirement of the absence of charge and color breaking (CCB) minima and unbounded from below (UFB) directions impose strong limits on the trilinear terms. In our analysis, we establish an upper bound for $k$ (remember that $a_0=k\, m_0$, \eq{eq:amismatch}) at the GUT scale and, after the running down to the EW scale, only points that satisfy the following relations are considered:

\bea
\mid (A_\ell)_{ii} \mid^2 & \leq & 3\, Y_{\ell_i}^2\, \left(\,m_{\ell_i}^2 \:+\: m_{e^c_i}^2 \:+\: m_{H_d}^2 \,\right),\, \label{eqn:colorcharge1}\\
\mid (A_\ell)_{ij} \mid^2 & \leq & Y_{\ell_k}^2\, \left(\,m_{\ell_i}^2 \:+\: m_{e^c_j}^2 \:+\: m_{H_d}^2 \,\right), \hspace{0.5cm} {\rm k\, =\, Max(i,j)} \label{eqn:colorcharge2}
\eea

As an application of these rules, we turn now to a phenomenological analyses using lepton flavor observables for three representative lepton flavor models found in the literature, based on the flavor groups $\mathcal{G}_f = $ $\Delta(27)$, $A_4$ and $S_3$.


\section{A $\Delta (27)$ Model} \label{del27model}

As a first example, we consider the flavor model of I. de Medeiros Varzielas. G. G. Ross and S. F. King in Ref.~\cite{deMedeirosVarzielas:2006fc}, where the continuum $SU(3)_f$ family symmetry of Ref.~\cite{deMedeirosVarzielas:2005ax}, already considered in our previous work to study the quark sector, was replaced by its discrete subgroup $\Delta(27)$. In this way the mechanism for obtaining the desired vacuum structure, which leads to  Tri-Bi-maximal (TB) mixing in the lepton sector through a type I see-saw mechanism, is considerably simplified.

$\Delta(27)$ is the simplest non-trivial group in the series $\Delta(3N^2)$, a discrete subgroup of $SU(3)$ that can be defined in terms of the semi-direct product $(Z_N\times Z'_N)\ltimes Z_3$. The elements of the group ($g$) can be written in terms of the generators of $Z_3$ ($a$, $a'$, $b$) as follows:
\beq
g \;=\; b^k\, a^m\, a'^n \hspace{0.5cm}{\rm for}\hspace{0.5cm} \mbox{k, m, n = 0, 1, 2}\, ,
\eeq
where the generators must satisfy
\bea
 a^3 \;=\; a'^3 & = & b^3 \;=\; e \hspace{1cm}\;,\;\hspace{1.6cm} a\,a' \;=\; a'\,a \nn \\
b\,a\,b^{-1} & = & a^{-1}a'^{-1} \hspace{1cm}\;,\;\hspace{1cm} b\,a'\,b^{-1} \;=\; a \, .
\eea
These conditions give rise to nine singlets and a triplet/anti-triplet representation. Table \ref{tab:D27pc} shows the particle content of the model: left-handed (LH) leptons transform as triplets $\bf 3$ whereas the right-handed (RH) fields transform as anti-triplets $\bar{\bf 3}$; the Higgs doublets are singlets under the group transformations and flavons, generically denoted as $\phi$, transform as triplet or anti-triplets.

\begin{table}[h!]
\centering 
\begin{tabular}{|c||c|c||c|c||c|c|c|c|c|}
\hline
~${\bf Field}$ & $\ell, \nu$ & $\ell^c,\nu^c$ & $H_{u,d}$ & $\Sigma$ &
$\phi_{123}$ & $\phi_1$ & $\bar\phi_3$ & $\bar\phi_{23}$ & $\bar\phi_{123}$\\ 
\hline
\hline
~$\Delta(27)$           & \bf 3         & \bf 3         & \bf 1 & \bf 1             & \bf 3 & \bf 3 & $\bar{\bf 3}$ & $\bar{\bf 3}$ & $\bar{\bf 3}$  \\
~$Z_2$    & 1 & 1 & 1 & 1 & 1 & -1 & -1 & -1 & -1 \\
~$U(1)_{FN}$    & 0    & 0  & 0 & 2  & -1 & -4 & 0 & -1 & 1 \\
~$U(1)_R$     & 1         & 1         & 0 & 0 & 0 & 0 & 0 & 0 & 0 \\ 
\hline
\end{tabular}
\caption{Transformation of the matter superfields under the $\Delta(27)$ family symmetries.}
\label{tab:D27pc}
\end{table}

Unlike the $SU(3)_f$ model, where the VEV of a triplet could be rotated to a single direction, the discrete non-Abelian symmetry leads to a finite number of candidate vacuum states. The obtained pattern for the VEVs is then given by \cite{deMedeirosVarzielas:2006fc}:
\bea
 \langle \bar\phi_3 \rangle^T & = &  \upsilon_3\,\left(\begin{array}{c} 0\\ 0\\ 1\end{array}\right)\hspace{.5cm},\hspace{.5cm}
 \langle \bar\phi_{23} \rangle^T \:=\:  \upsilon_{23}\,\left(\begin{array}{c} 0\\ -1\\ 1\end{array}\right) \,,\\
 \nn \\
 \langle \phi_{123} \rangle & \propto & \langle \bar\phi_{123} \rangle^T \:=\: 
  \upsilon_{123}\,\left(\begin{array}{c} 1\\ 1\\ 1 \end{array}\right) \hspace{.5cm},\hspace{.5cm} \langle \phi_1 \rangle \: 
 \propto \:  \upsilon_1\,\left(\begin{array}{c} 1\\ 0\\ 0 \end{array}\right)\,,
\eea

with $v_{123} \ll v_{23} \ll v_{3} \sim v_1$. 

The leading Yukawa terms responsible for the fermion masses in the $SU(3)_f$ model are still the dominant operators in this example although, beyond the LO, additional contributions enter in the superpotential. Its complete expression is \cite{deMedeirosVarzielas:2006fc}:
\bea
  {\cal W_\ell} & = & \frac{1}{M^2}\, (\ell\,\bar\phi_3)(\ell^c\,\bar\phi_3)\, H_d \;+\;  
  \frac{1}{M^2}\, (\ell\,\bar\phi_{23})(\ell^c\,\bar\phi_{123})\, H_d \;+\; \frac{1}
  {M^2}\,(\ell\,\bar\phi_{123})(\ell^c\bar\phi_{23})\, H_d \label{eqn:D27sp} \\
  & + & \frac{1}{M^3}\,(\ell\,\bar\phi_{23})\,(\ell^c\,\bar\phi_{23})\,\Sigma\,H_d \nn \\
  & + & \frac{1}{M^5}\, (\ell\,\bar\phi_{123})\,(\ell^c\,\bar\phi_{3})\,H_d\,\Sigma\,
  (\phi_{1}\,\bar\phi_{123})
    \;+\; \frac{1}{M^5}\, (\ell\,\bar\phi_{3})\,(\ell^c\,\bar\phi_{123})\,\Sigma\,H_d\,
    (\phi_{1}\,\bar\phi_{123}) \nn \\ 
  & + &\; \frac{1}{M^6}\,(\ell\,\bar\phi_{123})\,(\ell^c\,\bar\phi_{123})\,H_d\,
  (\phi_{123}\,\bar\phi_3)^2 \,.
  \nn \eea
After the flavor symmetry is broken, the Yukawa and Trilinear structures are given by:
\beq \label{eqn:D27YlAl}
 Y_\ell ~ \sim ~ y_\tau\left(\begin{array}{ccc}
             x_1\,\vep^8 &    -x_2\,\vep^3 & x_2\,\vep^3 \\
            -x_3\,\vep^3 &  3\,x_4\,\vep^2 & -3\,x_4\vep^2 \\
             x_3\,\vep^3 & -3\,x_4\,\vep^2 & x_5\,\alpha   
             \end{array} \right)
 \hspace{.5cm}, \hspace{.5cm}
 A_\ell ~ \sim ~ y_\tau\,a_{0} \left(\begin{array}{ccc}
                    13\,x_1\,\vep^8 &  -5\,x_2\,\vep^3 &   5\,x_2\,\vep^3 \\
                    -5\,x_3\,\vep^3 &  21\,x_4\,\vep^2 & -21\,x_4\,\vep^2 \\
                    5\,x_3\,\vep^3  & -21\,x_4\,\vep^2 &   5\,x_5\,\alpha   
                    \end{array} \right)  \, ,            
\eeq
where $x_i\sim {\cal O}(1)$, $\langle\Sigma\rangle/M_\ell\simeq-3$, $\upsilon_3/M_\ell=\alpha\simeq0.7$, $\upsilon_{123}/M_\ell\simeq \vep^2$ and the expansion parameter is given by $\vep=\upsilon_{23}/M_\ell\simeq0.15$. As stated before, $Y_\ell$ and $A_\ell$ are not simply proportional due to the mismatch caused by the different ways in which the spurion field can be attached to the Yukawa supergraphs in order to generate the Trilinear terms. Thus, from \eq{eq:amismatch}, the multiplicative factors in \eq{eqn:D27YlAl} are simply $2N +1$, with $N$ equal to the number of flavon insertions. For instance, in the case of $Y_{11} \propto \vep^8$, $N=6$ (see last line of the superpotential) and the proportionality factor would be $13$. Similarly, for $Y_{22} \propto 3\,\vep^2$, $N=3$ (second line of the superpotential) and $A_{ij}=7\,a_0\,Y_{ij}\propto 21\,\vep^2$.

Regarding the K\"ahler potential, it is important to stress here that in this model the $SU(2)_L$ doublet-messengers are assumed to be much heavier than their singlet counterpart. Because of that, corrections to the kinetic and soft terms for LH particles will be negligible and, therefore, the associated matrices can be taken as the identity matrix. In contrast, the LO K\"ahler potential for RH fields is:
\bea
  K_{\ell,R} & = & \ell^{c}\ell^{c\dagger} \;+\; \frac{1}{M^2}\left[( \ell^c \bar\phi_3)(\bar\phi_{3}^\dagger \ell^{c\dagger})\;+\; ( \ell^c \bar\phi_{23})(\bar\phi_{23}^\dagger \ell^{c\dagger}) \;+\; 
  ( \ell^c \bar\phi_{123})(\bar\phi_{123}^\dagger \ell^{c\dagger}) \right] \\
    & + & \frac{1}{M^3}\left[ ( \ell^c \bar\phi_{23})(\bar\phi_{123}^\dagger \ell^{c\dagger})\,\Sigma\, \;+\; {\rm h.c.} \right] \\
   & + &  \frac{1}{M^5} \left[ ( \ell^c \bar\phi_{123})\,(\bar\phi_{23}^\dagger \ell^{c\dagger})\,(\bar\phi_{3}\phi_1)\,\Sigma \;+\; {\rm 
    h.c.}\right] \, .\nn
\eea

Similarly, a mismatch between the soft-mass matrices and the K\"ahler metric will arise when considering the different ways in which $X X^\dagger$ can be coupled to the diagram, see Fig.~\ref{fig:kahlercorex1}. Once the flavons get their VEV, the K\"ahler function and soft-mass matrices can be written as:
\begin{equation} \label{eq:km2}
(K_{R})_{ij} = \left(\delta_{ij} \;+\; C_{R,\, ij}\, \right)\,, \qquad   (m^2_{R})_{ij} = m_0^2 \left(\delta_{ij} \;+\; B_{R,\, ij} \right) ,
\end{equation}

with $C_R$ and $B_R$ given by:
\beq
  C_R ~\sim~ \left(\begin{array}{ccc}
                             \vep^4 & -3\,(1+\alpha)\, \vep^3 & 3\,(1+\alpha)\, \vep^3 \\
              -3\,(1+\alpha)\, \vep^3 &          \quad \vep^2 &              -\vep^2 \\
            \quad3(1+\alpha)\, \vep^3 &               -\vep^2 &   \quad       \alpha^2 
             \end{array} \right)\, ,
\eeq

\beq \label{eqn:D27BLBR}
  B_R ~\sim~ \left(\begin{array}{ccc}
                    2\, \vep^4 & -3\,(3+5\,\alpha)\, \vep^3 & 3\,(3+5\,\alpha)\, \vep^3 \\
            -3\,(3+5\,\alpha)\, \vep^3 &   \quad 2\, \vep^2 & -2\, \vep^2 \\           
            \quad 3\,(3+5\,\alpha)\, \vep^3 &   -2\, \vep^2 & \quad    2\, \alpha^2              
             \end{array} \right) \,.
\eeq

\begin{figure}[h!]
  \centering
  \captionsetup{width=.9\linewidth}
  \vspace{-0.75cm}
  \includegraphics[scale=0.45]{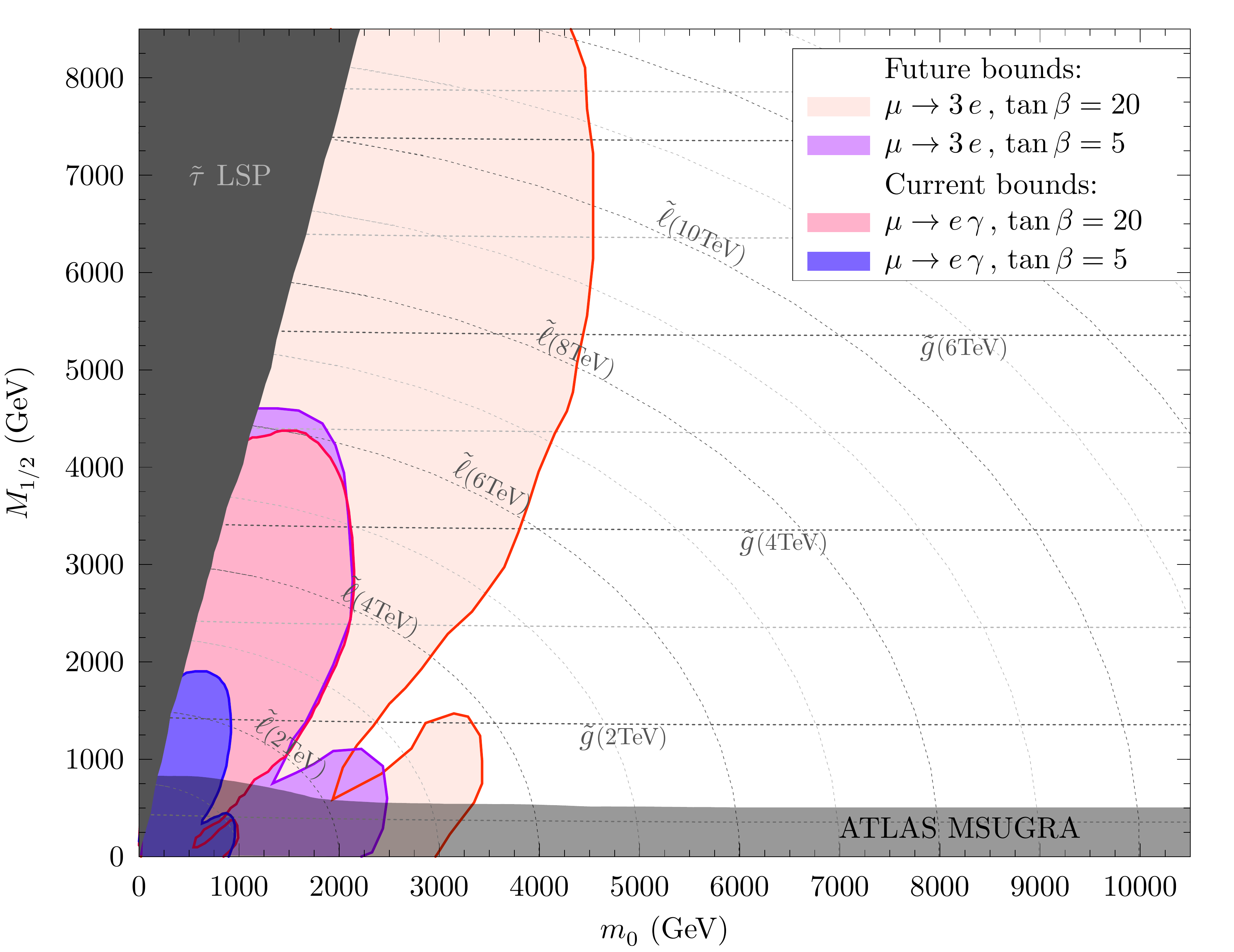}
  \caption{\label{fig:D27}Excluded regions due to $\mu \to e \gamma$  and $\mu\to eee$ for two reference values: $\tan\beta = 5$ (blue shapes)
and $\tan \beta = 20$ (red shapes). In the dark (blue and red) regions, we compare with current $\mu \to e \gamma$  bounds,
while in the light (blue and red) regions we compare with the expected  $\mu\to eee$ sensitivity in the near future.
Interestingly, even for present bounds, these results are competitive with mSUGRA ATLAS limits (gray area).}
\end{figure}

Again, the multiplicative factors in \eq{eqn:D27BLBR} can be easily computed from \eq{eq:kmismatch} just counting the number of flavon fields entering and leaving the diagram, without specifying the complete messenger spectrum of the UV theory.

With the structures of the Kinetic-mixing and Yukawa matrices known, the superfields must now be rotated twice: first, to the basis where canonical kinetic terms are recovered (canonical basis), and again, to the basis where the Yukawa couplings are diagonal (mass basis). Thus, the final matrices are:

\bea
  A_\ell & \longrightarrow & y_\tau\, a_{0}\, \left(\begin{array}{ccc}
 \cfrac{x_2 x_3}{x_4} \vep^4 & 2\,x_2\,\vep^3 & -2\,\cfrac{x_2}{x_5}\,\hat\alpha\;\vep^3 \\
               2\,x_2\,\vep^3 &    24\,x_4\,\vep^2 & -6\,x_4\,\hat\alpha\;\vep^2 \\
               -2\,x_2\,\vep^3 &  -6\,x_4\,\vep^2 & 5\,x_5\,\hat\alpha  
                             \end{array} \right)\, ,
\eea
\\
\bea
  m_{\ell,R}^2 & \longrightarrow &  m_{0}^2 \left(\begin{array}{ccc}                                    
             1 & -3\left(2+4\alpha\right) \vep^3 & 3\left(2+5\alpha-x_5\right)\,\hat\alpha\,\vep^3 \nn  \vspace{0.2cm} \\         
-3\left(2+4\alpha\right) \vep^3 & 1+\vep^2   & -\left(1+3\,\cfrac{x_4}{x_5}\,\alpha\right)\,\hat\alpha\,\vep^2 \nn \\
3\left(2+5\alpha-x_5\right)\,\hat\alpha\,\vep^3 & -\left(1+3\,\cfrac{x_4}{x_5}\,\alpha\right)\,\hat\alpha\,\vep^2 &  1+\hat\alpha^2\,\alpha^2
                                         \end{array} \right)\,,\\
  \eea
where $\hat\alpha\equiv1/\sqrt{1+\alpha^2}$.  The net effect of this series of rotations is the following: the canonical normalization makes the multiplicative factors of $B_R$ decrease by one unit, while having no impact on the Yukawa and Trilinear terms; the second rotation to the mass basis results in the reduction from $\vep^8\to\vep^4$ of $A_{\ell,11}$ and gives only additional small corrections to the elements of $B_R$. The matrix $U_\ell$ that performs the latter diagonalization gives only $\mathcal{O}(\vep^2)$ corrections to  $U_{\rm PMNS}=U_{\ell}^\dagger U_{\nu}$ so that it mantains the tri-bimaximal LO structure. As a consequence, this model cannot reproduce the experimental value of the reactor angle that would require $\sin\theta_{13}\propto \vep$ \footnote{After completion of this work, we came across the preprint~\cite{deMedeirosVarzielas:2017sdv}, where the authors succeed in obtaining a correct $\sin\theta_{13}$ in the context of a similar $\Delta (27)$ model.}. 

With these matrices, a combined fit to the latest experimental values for $U_{\rm PMNS}$ \cite{Patrignani:2016xqp}, excluding the 13 entry, and the Yukawas at the GUT scale \cite{Antusch:2013jca} is performed to fix the values of the $x_i$ coefficients. For $\vep=0.13$  these are reasonably $\mathcal{O}(1)$ coefficients, namely: ($x_1=1.0$, $x_2=1.2$, $x_3=1.$, $x_4=1.$, $x_5=1.7$). After substituting these values, the matrices must be run to the EW scale by means of the MSSM renormalization group equations (RGE), checked to satisfy the charge and color breaking relations, and compared to the most relevant flavor observables. Numerical calculations for the running, spectrum and low energy processes have been done with the Supersymmetric Phenomenology package (SPheno) \cite{Porod:2011nf, Porod:2003um}. The resulting plot is Fig.~\ref{fig:D27}.

As shown in Fig.~\ref{fig:D27}, the most restrictive constraints come from the flavor violating decays $\mu \to e \gamma$ and $\mu \to e e e$. In the plot, the colored shapes represent the parameter regions where the analyzed model would dissagree with current and future bounds in Table \ref{tab:LFVbounds}. As the results strongly depend on $\tan\beta$, two reference values of $\tan\beta$ has been considered that is $\tan\beta = 5$, blue (darker) regions, and $\tan\beta= 20$, red (lighter) regions. It can be observed that, for both values of $\tan\beta$, the obtained bounds are competitive with mSUGRA ATLAS limits, even just considering present $\mu \to e \gamma$ experimental limits. On top of that, if the Mu3e experiment reaches the expected precision finding no sign of the $\mu \to e e e$ process, the parameter space of the model will turn out to be significantly constrained.

These results are in good agreement with those obtained with the mass insertion approximation (MIA) \cite{Hall:1985dx,Gabbiani:1988rb,Gabbiani:1996hi,Paradisi:2005fk,Ciuchini:2007ha}, which provides a simplified description of the phenomenology. As discussed in \cite{Ciuchini:2007ha,Calibbi:2008qt,Calibbi:2009ja}, in the absence of off-diagonal $\delta_{LL}$ insertions, the main effects come from the RR sector. This sector suffers from a characteristic cancellation among the two $\tan\beta$-enhanced dominant contributions: the one due to the pure bino term (with internal chirality flip and a flavor-conserving $\delta_{LR}$ mass insertion) and another from the bino-higgsino exchange. This destructive interference can be easily recognized in Fig.~\ref{fig:D27}. Moreover, these contributions require a bino mass insertion, $M_1$, so, as we see in the figure, the bound practically disappears for small values of $M_{1/2}$.


\section{An $A_4$ Model}  
\label{sec:a4model}

As a second example, we consider a model belonging to perhaps the most popular class of models based on discrete flavors groups, those with $\mathcal{G}_f = A_4$. This is the discrete group of even permutations of 4 objects; it contains 12 elements and has four inequivalent irreducible representations: three singlets $\lbrace {\bf 1,\, 1',\, 1''}\rbrace$ and a triplet $\bf 3$. It is specially interesting because it is the minimal non-Abelian group containing a triplet representation. We refer to Appendix \ref{app:A4group} for a detailed description of the group, including the associated multiplication rules.

Flavor models based on an $A_4$ symmetry \cite{Altarelli:2005yx, Altarelli:2009kr, Altarelli:2005yp, Ma:2001dn, Babu:2002dz, Ma:2004zv, He:2006dk, Babu:2005se, Zee:2005ut, Ma:2005qf, King:2006np, Altarelli:2006kg, Hirsch:2007kh, Bazzocchi:2007na, Honda:2008rs, Bazzocchi:2008rz, Hirsch:2008rp, Lin:2008aj, Csaki:2008qq, Feruglio:2008ht, Morisi:2007ft, Hirsch:2005mc, Hirsch:2003dr, Chen:2009um, Hirsch:2003xx} have been an attractive option for describing the lepton sector due to their simplicity and economical structure in reproducing the well-known TB-mixing pattern at leading order (LO). Although this scheme predicts a vanishing reactor angle, currently excluded by data \cite{Abe:2014bwa, An:2015rpe, RENO:2015ksa}, variations of these models \cite{Lin:2009bw, Cooper:2012wf, Hernandez:2012ra, King:2011zj, Zheng:2011uz, Ma:2011yi, Rodejohann:2011uz, Ahn:2011if, Kumar:2011vf, Gupta:2011ct, Branco:2012vs, Ahn:2012tv, Altarelli:2008bg} may still accommodate an adequate $\theta_{13}$, once higher order corrections to masses and mixings are taken into account.

\begin{table}[t!]
\centering 
\begin{tabular}{|c||c|c|c|c|c||c|c||c|c|c|c|c|}
\hline
~${\bf Field}$ & $\nu^c$ & $\ell$ & $e^c$ & $\mu^c$ & $\tau^c$ & $H_d$ & $H_u$ & $\phi_S$  & $\phi_T$ & $\xi$ & $\xi'$ & $\xi'^\dagger $  \\ 
\hline
\hline
~$A_4$    & \bf 3 & \bf 3 & \bf 1 & \bf 1 & \bf 1 & \bf 1 & \bf 1 & \bf 3 & \bf 3 & \bf 1 & \bf 1$'$  & \bf 1$''$  \\
~$Z_4$    &    -1 &     i &     1 &     i &    -1 &     1 &     i &     1 &     i  &     1 &   i &   -i  \\
~$U(1)_R$ &     1 &     1 &     1 &     1 &      1 &     0 &    0 &     0 &     0  &     0 &    0 &  0  \\
\hline
\end{tabular}
\caption{Transformation of the matter and flavon superfields under the flavor symmetry $\mathcal{G}_f=A_4\times Z_4$, for non trivial cases the correspondent daggered fields are also specified.}
\label{tab:A4pc}
\end{table}

Here, we analyze the $A_4$ Altarelli-Meloni model of Ref.~\cite{Altarelli:2009kr}, which can be seen as a simplest $A_4$ model in the sense that it is able to generate an appropriate charged-lepton hierarchy between generations without requiring an extra $U(1)_{FN}$ symmetry. The complete flavor symmetry of the model is $\mathcal{G}_f=A_4\times Z_4$ with an additional $U(1)_R$ symmetry related to R-parity. Table \ref{tab:A4pc} shows the symmetry assignments for leptons, electroweak Higgs doublets and flavons. In particular, the three generations of left-handed lepton doublets $\ell$ and the right-handed neutrino $\nu^c$ are ascribed to triplet representations while the right-handed charged leptons $e^c$, $\mu^c$, $\tau^c$, together with the two Higgs doublets $H_{u,d}$, transform in the trivial singlet representation. Beyond the MSSM fields, the model contains  the flavons that transform as singlets or triplets.

The vacuum alignment in this model responsible for the symmetry breaking~\cite{Altarelli:2009kr} is given by
\bea 
\label{eqn:A4vevs}
 	\langle\phi_T\rangle \,\propto\,  \upsilon_T\, \left(\begin{array}{c} \delta  \hat\upsilon_{T1} \\ 1+\delta  \hat \upsilon_{T2}\\\delta  \hat\upsilon_{T3}\end{array} \right) \hspace{0.5cm}&,&\hspace{1cm} \nn
	\langle\phi_S\rangle \,\propto\, \upsilon_S\, \left(\begin{array}{c} 1+\delta  \hat v_{S} \\ 1+\delta \hat v_{S}\\1+\delta  \hat v_{S} \end{array} \right) \, ,\\\\ \nn
	\langle\xi\rangle \,\propto\, \upsilon_\xi \hspace{1.5cm}&,&\hspace{1cm}
 	\langle\xi'\rangle \,\propto\, \upsilon_\xi '\,(1+\delta \hat \upsilon'_{\xi})\, ,
\eea

where $\delta \hat \upsilon_{i}= \delta \upsilon_{i}/M$, $\upsilon_T/M\sim \upsilon_\xi '/M\sim\vep$ and $\upsilon_S/M\sim \upsilon_\xi /M\sim \delta \upsilon_{i}/M\sim \vep'$. The shift in the VEVs, denoted as $\delta\upsilon_{i}$, account for NLO corrections arising from higher-order operators in the driving superpotential. A similar order of magnitude is expected for $\vep$ and $\vep'$, although a moderate hierarchy can be tolerated among them. 

The LO effective superpotential contains the following operators
\bea
\label{eqn:A4W} 
  {\cal W}_\ell & = & \frac{1}{M}\, \tau^c(\ell \phi_T )\, H_d \nn\\
             		& + & \frac{1}{M^2}\, \mu^c\left[(\ell \phi_T^2)\, \:+\: (\ell \phi_T)''\xi'\, \right]\,H_{d}  \\
             		& + &  \frac{1}{M^3}\, e^c\left[ (\ell \phi_T^3) \nn \:+\:\, (\ell \phi_T^2)''\xi'\:+\:\, (\ell \phi_T)'\xi'^2\, \right] H_d \, ,\nn       
\eea
where the brackets stand for each possible product combination of the fields inside. It is easy to see that, replacing Eqs.~(\ref{eqn:A4vevs}) into \eq{eqn:A4W} with {\small $\delta \upsilon_{i}=0$}, the vacuum configuration leads to diagonal and hierarchical Yukawas in the charged-lepton sector. Off-diagonal entries in the Yukawa matrix derive from considering the shifted VEVs ({\small $\delta\upsilon_i\neq 0$}) in the LO superpotential and higher-order operators obtained by the insertion of $\phi_S$ and $\xi'$~\cite{Altarelli:2009kr}. Taking into account the charges of Table \ref{tab:A4pc}, the correction to the LO superpotential would be:
\bea
\label{eqn:A4dW}
 {\cal \delta W}_\ell & = &  \frac{1}{M^2}\, \tau^c\left[(\ell \phi_T \phi_S)\:+\: (\ell \phi_S)''{\xi'}\, \right]\,H_{d}\nn\\
             			& + & \frac{1}{M^3}\, \mu^c\left[(\ell \phi_T^2 \phi_S)\:+\:(\ell \phi_T \phi_S)''\xi' \:+\: (\ell\phi_S)'\xi'^2 \,\right]\,H_{d} \\       
				& + &  \frac{1}{M^4}\, e^c\left[(\ell \phi_T^3 \phi_S)\nn \:+\: (\ell \phi_T^2\phi_S)''\xi'\:+\:(\ell \phi_T\phi_S)'\xi'^2 \:+\:\,(\ell \phi_S)\xi'^3 \,\right]\, H_d\,. \nn         
\eea 
As can be seen in \eq{eqn:A4YlAl}, these contributions result in non-vanishing off-diagonal entries of the same order of the diagonal term in each row multiplied by $\vep'$:
\beq 
\label{eqn:A4YlAl}
 Y_\ell ~ \sim ~ \left(\begin{array}{ccc}
               x_1\,\vep^3      &  x_2\,\vep^3\vep' & x_3\,\vep^3\vep' \\
               x_4\,\vep^2\vep' &  x_5\,\vep^2      & x_6\,\vep^2\vep' \\
               x_7\,\vep\,\vep' &  x_8\,\vep\,\vep' & x_9\,\vep\,   
               \end{array} \right)
 \hspace{0.5cm} , \hspace{0.5cm}
 A_\ell ~ \sim ~  a_{0}\: \left(\begin{array}{ccc}
               7\,x_1\,\vep^3      &  9\,x_2\,\vep^3\vep' & 9\,x_3\,\vep^3\vep' \\
               7\,x_4\,\vep^2\vep' &  5\,x_5\,\vep^2      & 7\,x_6\,\vep^2\vep' \\
               5\,x_7\,\vep\,\vep' &  5\,x_8\,\vep\,\vep' & 3\,x_9\,\vep\,   
               \end{array} \right)         
\eeq

with $x_i\, \sim {\cal O}(1)$ generic order one coefficients. Again, $Y_\ell$ and $A_\ell$ are not proportional and the multiplicative factors in the Trilinears can be computed with \eq{eq:amismatch} considering $N$ equal to the power associated to $\vep$ and/or $\vep'$ in the correspondent Yukawa element. 

The LO K\"ahler potential for left-handed (LH) fields is given by: 
\bea \label{eqn:A4KL}
  K_{\ell,\, L} & = & \ell\,\ell^\dagger \;+\; \frac{1}{M^2}\left[(\ell\,\ell^\dagger\,\phi_S\,\phi_S^\dagger) \;+\; (\ell\,\ell^\dagger\,\phi_S)\,\xi^\dagger\right] \;+\;  {\rm h.c.}\,,
\eea

whereas the right-handed (RH) K\"ahler potential would be:
\bea \label{eqn:A4KR}
  K_{\ell,\, R} & = & e^c e^{c\dagger}\;+\; \mu^c \mu^{c\dagger}\;+\; \tau^c \tau^{c\dagger}\;+\; \nn\\
  & + & \frac{1}{M^2}\, \left[e^c (\phi_T\phi_S^\dagger )\mu^{c\dagger} \;+\; \mu^c(\phi_T\phi_S^\dagger ) \tau^{c\dagger}\, \right] \\
  & + & \frac{1}{M^3}\; e^c \left[\, (\phi_S\phi_T^{\dagger\, 2}) \:+\: (\phi_S\phi_T^{\dagger})'\xi'^\dagger \:+\: {\rm h.c.} \right]\,\tau^{c\dagger} \;+\; {\rm h.c.}\, \nn,
\eea

Once the flavons have been integrated out, the K\"ahler function and soft-mass matrices for both LH- and RH-fields can be written as in \eq{eq:km2} with $C_{L(R)}$ and $B_{L(R)}$:

\bea
  C_L & \sim~ \left(\begin{array}{ccc}
        	  	   \vep^2+\vep'^2 &        \vep'^2 & \vep'^2 \\
                      \vep'^2 & \vep^2+\vep'^2 & \vep'^2 \\
                      \vep'^2 &        \vep'^2 & \vep^2+\vep'^2
             \end{array} \right)
             \hspace{0.8cm} , \hspace{0.8cm}
 C_R ~~\sim~ \left(\begin{array}{ccc}
               		\vep^2+\vep'^2 & \vep\,\vep'     & \vep^2\vep' \\
                      \vep\,\vep'  & \vep^2+\vep'^2  & \vep\,\vep' \\
                	    	\vep^2\vep'    & \vep\,\vep' 	& \vep^2+\vep'^2
             \end{array} \right)   \, ,\nn\\
\eea
\bea             
  \hspace{0.15cm} B_L & \sim~  2\; \left(\begin{array}{ccc}
                  \vep^2+\vep'^2 	&        \vep'^2 & \vep'^2 \\
                         \vep'^2 	& \vep^2+\vep'^2 & \vep'^2 \\
                         \vep'^2 	&        \vep'^2 & \vep^2+\vep'^2
             \end{array} \right)
             \hspace{0.6cm} , \hspace{0.8cm}
  B_R ~~\sim~ 2\; \left(\begin{array}{ccc}
               		 \vep^2+\vep'^2 	& \vep\,\vep'    & \frac{3}{2}\,\vep^2\vep' \\
                        \vep\,\vep'  & \vep^2+\vep'^2 & \vep\,\vep' \\
            \frac{3}{2}\,\vep^2\vep' &  \vep\,\vep' 	 & \vep^2+\vep'^2
             \end{array} \right)\,.\nn\\     
 \label{eqn:A4BLBR}
\eea

Again, the multiplicative factors in \eq{eqn:A4BLBR} can be easily figured out from \eq{eq:kmismatch} by just computing the number of flavon fields entering and leaving the diagram. Then, we perform the two rotations to the canonical and the mass basis that result in the following rotated matrices
\bea
A_\ell & \longrightarrow & a_{0}\: \left(\begin{array}{ccc}
               7\,x_1\,\vep^3    &  \left(4\,x_2+2\cfrac{x_1 x_4}{x_5}\right)\,\vep^3\vep' & \left(6\,x_3+4\cfrac{x_1 x_7}{x_9}\right)\,\vep^3\vep' \\
               2\,x_4\,\vep^2\vep' &  5\,x_5\,\vep^2        & \left(4\,x_6+2\cfrac{x_5x_8}{x_9}\right)\,\vep^2\vep' \\
               2\,x_7\,\vep\,\vep'     &  2\,x_8\,\vep\,\vep'       & 3\,x_9\,\vep\,   
               \end{array} \right)    \,,\\ 
\nn \\ 
\nn \\               
m_{\ell,L}^2 & \longrightarrow & m_{0}^2\: \left(\begin{array}{ccc}
               1+\vep^2+\vep'^2    &  \vep'^2 & \vep'^2  \\
               \vep'^2  &  1+\vep^2+\vep'^2 & \vep'^2  \\
               \vep'^2     &  \vep'^2    & 1+\vep^2+\vep'^2 
               \end{array} \right)    \,,\\   
\nn \\  
\nn \\             
m_{\ell,R}^2 & \longrightarrow & m_{0}^2\: \left(\begin{array}{ccc}
               1+\vep^2+\vep'^2    &  \vep\,\vep' & 2\,\vep^2\vep' +\left(\cfrac{x_4}{x_5}-\cfrac{x_8}{x_9}\right)\vep\,\vep'^2  \\
               \vep\,\vep'  &  1+\vep^2+\vep'^2 &\vep\,\vep'  \\
              2\,\vep^2\vep' +\left(\cfrac{x_4}{x_5}-\cfrac{x_8}{x_9}\right)\vep\,\vep'^2  &  \vep\,\vep'    & 1+\vep^2+\vep'^2    
               \end{array} \right)    \,.\nn\\
\eea

We find that the dominant structures of the matrices remain unaltered, the coefficients receiving only small corrections. In this case, the Yukawa rotation matrix $U_\ell$ gives rise to an $\mathcal{O}(\vep')$ correction to the 13 entry of the PMNS matrix, such that the model can reproduce the experimental magnitude of $\sin\theta_{13}$.  This imposes $\vep' \sim 0.1$ while the value of $\vep$ is fixed by the Yukawa hierarchy. Note that the off diagonal entries in the soft mass matrices arise at order $\vep'^2$. 

The ${\cal O}(1)$ coefficients $x_i$ are determined by the combined fit of the experimental values of $U_{\rm PMNS}$ \cite{Patrignani:2016xqp} and the Yukawas at the GUT scale \cite{Antusch:2013jca}. For $\tan\beta\,=\,5$ and $(\vep,\vep')=(0.04\,,\,0.08)$ we obtain: ($x_1=0.7$, $x_2=1.0$, $x_3=-1.0$, $x_4=1.6$, $x_5=5.3$, $x_6=0.99$, $x_7=4.0$, $x_8=5.4$, $x_9=3.6$); whereas for $\tan\beta \,=\,20$ and $(\vep,\vep')=(0.02\,,\,0.06)$, ($x_1=1.3$, $x_2=1.0$, $x_3=0.99$, $x_4=1.8$, $x_5=5.3$, $x_6=0.99$, $x_7=4.4$, $x_8=0.81$, $x_9=1.8$).
\\
\begin{figure}[h!]
  \centering
  \captionsetup{width=.9\linewidth}
  \vspace{-0.75cm}
  \includegraphics[scale=0.45]{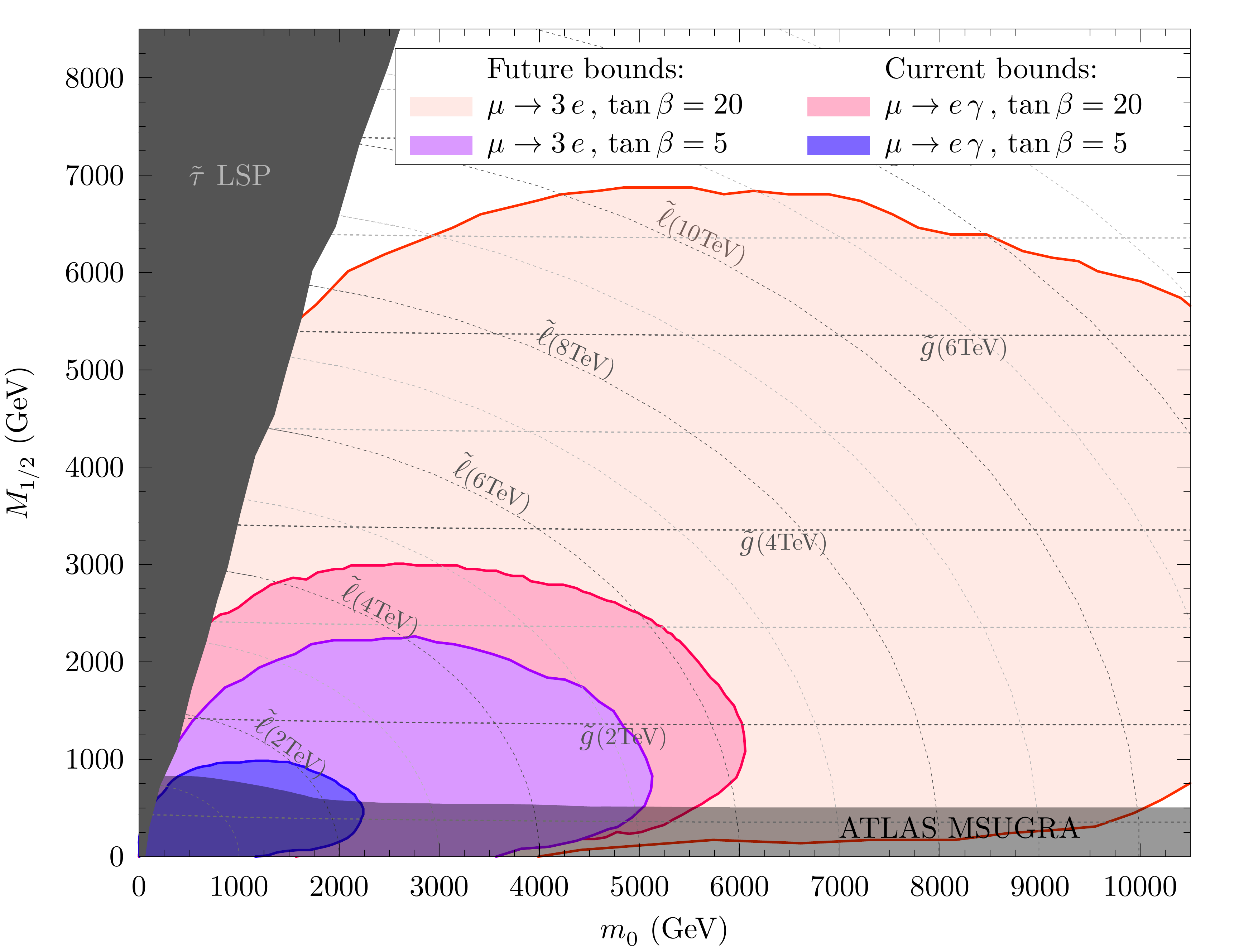}
  \caption{Excluded regions due to $\mu \to e \gamma$  and $\mu\to eee$ for two reference values: $\tan\beta = 5$ (blue shapes)
and $\tan \beta = 20$ (red shapes). In the dark (blue and red) regions, we compare with current $\mu \to e \gamma$  bounds,
while in the light (blue and red) regions we compare with the expected $\mu\to eee$ sensitivity in the near future.
As before, these results are competitive with mSUGRA ATLAS limits (gray area).}
  \label{fig:A4currents}
\end{figure}

After RGE evolving the matrices to the SUSY scale with SPheno, checking the charge and color breaking relations, and calculating the low-energy observables, the constraints on the model are shown in Fig.~\ref{fig:A4currents} for $\tan\beta = 5$, blue (dark) region, and $\tan\beta =20$, red (light) region. As expected, the most restrictive constraints come from the flavor violating decays $\mu \to e \gamma$ and $\mu \to e e e$. Current limits of the first process are competitive with present ATLAS bounds whereas future limits for $\mu \to e e e$ will allow us to either discover SUSY or to constraint a considerable part of the parameter space if no signal is measured.

In contrast with the previous example, no cancellation is observed here. This is because, in this model, the dominant effect comes from the LL mass insertion and, therefore, the two $\tan\beta$-enhanced terms have the same sign. A detailed discussion of these effects can be found in \cite{Ciuchini:2007ha}. We see that present and future LFV constraints are able to explore large values of $m_0$ and $M_{1/2}$ in these models, well beyond the LHC reach. 


\section{An $S_3$ Model}  \label{s3model}

Finally, another interesting and minimal group of models are those based on the symmetry group $S_3$ \cite{Haba:2005ds, Morisi:2005fy, Chen:2004rr, Xing:2010iu, Grimus:2005mu, Mohapatra:2006pu, Feruglio:2007hi, Ma:1991eg} defined as the group of all possible permutations among $3$ objects, containing only $6$ group elements. The number of irreducible representations is $3$, which includes two singlets, $\lbrace{\rm \bf 1},\, {\rm \bf 1'}\rbrace$, and a doublet, $\rm \bf 2$. The detailed description of the group can be found in Appendix \ref{app:S3group}. 
The example that we consider here is the model of D. Meloni in Ref.\cite{Meloni:2012ci}, which generates a PMNS LO-structure compatible with the TB-mixing together with a relatively large reactor angle and a good description of the quark sector. The full flavor symmetry of the model is ${\cal G}_f=S_3\times Z_6\times Z_3$ with and additional $U(1)_R$ continuous symmetry which will eventually break down to R-parity due to small SUSY breaking effects. 
\begin{table}[h!]
\centering 
\begin{tabular}{|c||c|c|c|c|c|c||c||c|c|c|c|c|}
\hline
~${\bf Field}$ & $\nu^c$ & $\nu^c_3$ & $e$ & $e^c$ & $\ell$ & $\ell^c$ & $H_{u,d}$ & $\phi$ & $\chi$ & $\xi$ & $\chi'$ & $\chi'^{\dagger}$  \\ 
\hline
\hline
~$S_3$  & \bf 2 &${\bf 1'}$ &\bf 1 &\bf 1 &\bf 2 &\bf 2 &\bf 1 &\bf 2 &\bf 1  &\bf 2 &${\bf 1'}$  &${\bf 1'}$ \\
~$Z_6$     &   $\omega$ &    $\omega$ &     1 & $\omega^3$ & $\omega^5$ & $\omega^3$ &     1 & $\omega^4$ &$\omega^4$ & $\omega^4$ &$\omega^5$ &$\omega^{-5}$  \\
~$Z_3$     &          1 &           1 &     1 &   $\omega$ &          1 & $\omega^2$ &     1 &   $\omega$ &      
 $\omega$  &          1 &           1&           1  \\
~$U(1)_R$  &          1 &           1 &     1 &          1 &          1 &          1 &     0 &          0 &     
        0  &          0 &           0&           0 \\
\hline
\end{tabular}
\caption{Transformation of the matter superfields under the $S_3$ family symmetries.}
\label{tab:S3pc}
\end{table}

Table \ref{tab:S3pc} shows the complete spectrum for this model. As can be seen, the $SU(2)_L$ doublets and singlets of the second and third generations are arranged in two $S_3$ doublets, $\ell$ and $\ell^c$:
\beq
  \ell~=~\left(\begin{array}{c}
                \tau \\
                \mu \end{array}\right) \hspace{1.cm},\hspace{1.cm}
  \ell^c~=~\left(\begin{array}{c}
               \mu^c \\
              \tau^c \end{array}\right) \,,
\eeq

whereas the electron fields are assigned to the real singlets, $e$ and $e^c$. The electron and muon Majorana neutrinos are grouped in a doublet, $\nu^c$, while the tau right-handed neutrino transforms as the pseudosinglet representation, $\nu_3^c$. The minimization of the driving superpotential in the exact SUSY limit generates the desired alignment for the vacuum structure \cite{Meloni:2012ci}:
\beq \label{eqn:S3vac1}
  \langle\phi\rangle \,\propto\, \upsilon_\phi\, \left(\begin{array}{c} 1 \\ 1 \end{array} 
  \right) \hspace{0.5cm},\hspace{0.5cm}\langle\xi\rangle \,\propto\, \upsilon_\xi\, \left(\begin{array}{c} 
  \delta \hat{\upsilon}_\xi \\ 1 \end{array} \right)\,,\nn
  \eeq

\beq \label{eqn:S3vac2}
\langle\chi\rangle \,\propto\, \upsilon_\chi \hspace{1cm},\hspace{1cm}
  \langle\chi'\rangle \,\propto\, \upsilon_\chi'\, ,
\eeq

where $\delta \hat \upsilon_{\xi}= \delta \upsilon_{\xi}/M$, $\upsilon_\phi/M\sim \upsilon_\chi/M\sim\vep$ and $\upsilon_\xi/M\sim \upsilon_{\chi'}/M\sim \delta \upsilon_{\xi}/M\sim \vep'$. 

At LO, only the muon and tau masses are generated by operators involving one and two flavon insertions while the electron remains massless. To obtain its mass, operators with up to 5-flavon insertions must be considered. The dominant terms are given by the following contributions:
\bea \label{eqn:S3Weff}
  {\cal W}_\ell & = & \frac{1}{M}\, [\, (\ell^{c} \ell\,\phi)\;+\; (\ell^{c} \ell)\,\chi \,] \,H_{d} \nn\\
  			 & + & \hspace{-0.1cm} \frac{1}{M^{2}}\,(\ell^{c} \ell \phi)'\, \chi'\, H_{d} \nn \\
  			 & + & \hspace{-0.15cm} \frac{1}{M^4}\,e^c\,[(\ell\,\xi^2)\chi^2 \;+\; (\ell\,\phi\xi^2)\chi \;+\; (\ell\, \phi^2\xi^2) \,]\, H_d \nn \\
  			 & + & \hspace{-0.15cm} \frac{1}{M^5}\, e^ce\, [\,(\phi\,\xi^2)'\chi'\chi \;+\; (\phi^2\xi^2)'\chi'\,]\, H_d
\eea
In the vacuum alignment configuration, Eqs.~(\ref{eqn:S3vac1}) and (\ref{eqn:S3vac2}), the resulting effective Yukawa and Trilinear matrices are:
\beq
  Y_l ~ \sim ~ \left(\begin{array}{rrr}
             x_1\,\vep^{2}\,\vep'^{3} & x_2\,\vep\,\vep' & -x_2\,\vep\,\vep' \\
             x_3\,\vep^{2}\,\vep'^{2} & x_4\,\vep        &  x_5\,\vep \\
             x_6\,\vep^{2}\,\vep'^{2} & x_5\,\vep        &   x_4\,\vep   
             \end{array} \right)
  \hspace{0.7cm},\hspace{0.7cm}
  A_l ~ \sim ~ A_{0} \left(\begin{array}{rrr}
             11\,x_{1}\,\vep^{2}\,\vep'^{3} & 5\,x_2\,\vep\,\vep' & -5\,x_2\,\vep\,\vep' \\
              9\,x_3\,\vep^{2}\,\vep'^{2} &  3\,x_4\,\vep &    3\,x_5\,\vep  \\
              9\,x_6\,\vep^{2}\,\vep'^{2} &    3\,x_5\,\vep &   3\,x_4\,\vep    
             \end{array} \right)\, ,
\eeq

where the proportionality factor between each Yukawa and Trilinear term is given again by \eq{eq:amismatch}, with $N$ equal to the total power of $\vep$ and $\vep'$.

The LO contributions in the K\"ahler potential for LH- and RH-fields are given by:
\bea
  K_{\ell,L} & = & \ell\,{\ell}^\dagger \;+\; e\,e^\dagger \;+\; \frac{1}{M^2}\, \bigg[\, \left(\ell\,\ell^\dagger \phi \phi^\dagger\right) \;+\; \left(\ell\,\ell^\dagger \phi \right)\chi^\dagger \;+\; \chi'\left(\ell\,\xi^\dagger \right)'e^\dagger \,+\, {\rm h.c.} \,\bigg] \;+\; {\rm h.c.} \\
  \nn \\
  K_{\ell,R} & = & \ell^c \ell^{c\dagger} \;+\; e^c e^{c\dagger} \;+\; \frac{1}{M^2}\, \bigg[\, \left(\ell^c\ell^{c\dagger}\phi\phi^\dagger\right)  \;+\; \left(\, \ell^c\ell^{c\dagger}\phi\right)\chi^\dagger  \;+\; \left(\ell^c\xi\phi^\dagger\right)e^{c\dagger} \,+\, {\rm h.c.} \,\bigg] \;+\; {\rm h.c.} \nn\\
\eea

Once the flavor symmetry is broken, the K\"ahler metric and soft-mass matrices can be written in terms of $C_{L,R}$ and $B_{L,R}$ as in \eq{eq:km2}, with:
\begin{align}
  C_L & \sim~ \left(\begin{array}{ccc}
               \vep^2+\vep'^2 &        \vep'^2 & \vep^2\vep' \\
                      \vep'^2 & \vep^2+\vep'^2 &   \vep^2 \\
                  \vep^2\vep' &         \vep^2 & \vep^2+\vep'^2
             \end{array} \right)\,,
  &C_R ~\sim~ \left(\begin{array}{ccc}
               \vep^2+\vep'^2 &      \vep\vep' & \vep\vep' \\
                    \vep\vep' & \vep^2+\vep'^2 &   \vep^2 \\
                    \vep\vep' &        \vep^2 & \vep^2+\vep'^2
             \end{array} \right) \,,          \\
 \nn \\
  B_L & \sim~ 2\,\left(\begin{array}{ccc}
                  \vep^2+\vep'^2 &        \vep'^2 & \frac{3}{2}\vep^2\vep' \\
                         \vep'^2 & \vep^2+\vep'^2 & \vep^2 \\
          \frac{3}{2}\vep^2\vep' &         \vep^2 & \vep^2+\vep'^2
             \end{array} \right)\,,
  & \hspace{1.5cm} B_R ~\sim~ 2\, \left(\begin{array}{ccc}
               \vep^2+\vep'^2 &      \vep\vep' & \vep\vep' \\
                    \vep\vep' & \vep^2+\vep'^2 &   \vep^2 \\
                    \vep\vep' &        \vep^2 & \vep^2+\vep'^2
             \end{array} \right) \, .  \label{eqn:S3bLbR}    
\end{align}

After canonical normalization and diagonalization of the Yukawa matrices, the soft terms in the mass basis are:
\bea
A_\ell & \longrightarrow & a_{0}\: \left(\begin{array}{ccc}
            11\,x_1\,\vep^2\vep'^3 & \left(-\cfrac{5}{\sqrt{2}}\,x_2+\cfrac{3\sqrt{2}x_2x_5}{x_4+x_5}\right)\,\vep^3\vep' & -2\sqrt{2}\,x_2\,\vep^3\vep' \\
             \cfrac{9}{\sqrt{2}}(x_6+x_3)\,\vep^2\vep'^2 & 3\,(x_5-x_4)\vep & -3\,x_5\,\vep^3 \\
             \cfrac{9}{\sqrt{2}}\,(x_6-x_3)\,\vep^2\vep'^2 & -3\,x_5\,\vep^3 & -3\,(x_5+x_4)\vep  
               \end{array} \right)    \,,\\ 
\nn \\ 
\nn \\             
m_{\ell,L}^2 & \longrightarrow & m_{0}^2\: \left(\begin{array}{ccc}
            1+\vep^2+\vep'^2       & \cfrac{1}{\sqrt{2}}\,\vep'^2 & -\cfrac{1}{\sqrt{2}}\,\vep'^2 \\
         \cfrac{1}{\sqrt{2}}\,\vep'^2 & 1+2\,\vep^2+\vep'^2           & 3\,\vep^2\vep'^2  \\
        -\cfrac{1}{\sqrt{2}}\,\vep'^2 & 3\,\vep^2\vep'^2            & 1+\vep'^2
               \end{array} \right)    \,,\\   
\nn \\    
\nn \\         
m_{\ell,R}^2 & \longrightarrow & m_{0}^2\: \left(\begin{array}{ccc}
         1+\vep^2+\vep'^2 & \sqrt{2}\,\vep\,\vep'    & {\cal O}(\vep^3\vep'^3)  \\
     \sqrt{2}\,\vep\,\vep'  & 1+2\,\vep^2+\vep'^2        & {\cal O}(\vep^4\vep'^2)  \\
    {\cal O}(\vep^3\vep'^3) &  {\cal O}(\vep^4\vep'^2) & 1+\vep'^2
               \end{array} \right)    \,.
\eea
The ${\cal O}(1)$ coefficients $x_i$ are set so that the experimental values for $U_{\rm PMNS}$ \cite{Patrignani:2016xqp} and the Yukawa couplings at the GUT scale \cite{Antusch:2013jca} are reproduced. Both for $\tan\beta \,=\,5$ with $(\vep\,,\,\vep')=(0.08\,,\,0.08)$, and $\tan\beta \,=\,20$ with $(\vep\,,\,\vep')=(0.1,0.08)$ we obtain almost the same coefficients, that is ($x_1=3.$, $x_2=1.6$, $x_3=2.3$, $x_4=0.6$, $x_6=2.2$).

\begin{figure}[h!]
  \centering
  \captionsetup{width=.9\linewidth}
  \vspace{-0.75cm}
  \includegraphics[scale=0.45]{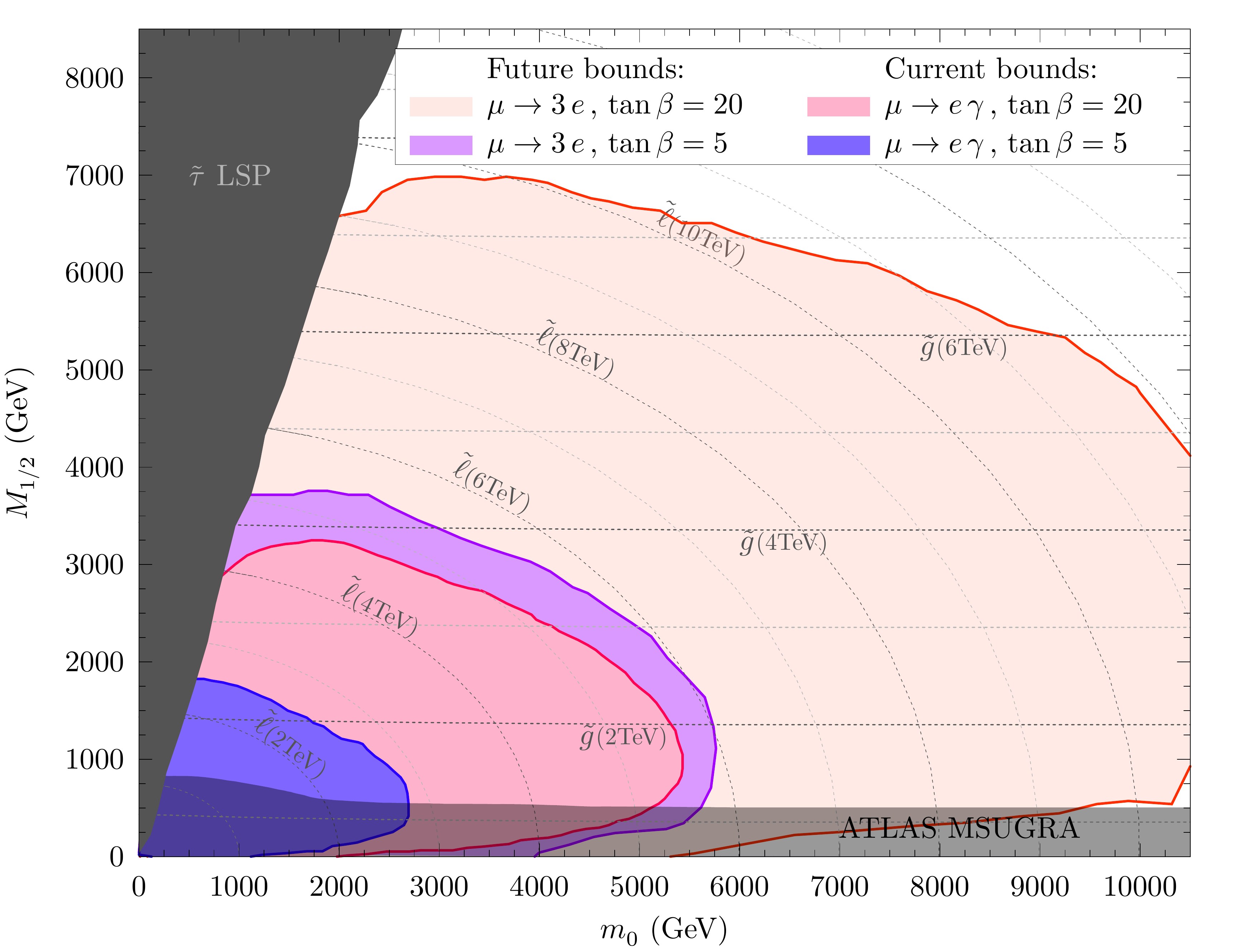}
  \caption{Excluded regions due to $\mu \to e \gamma$  and $\mu\to eee$ for two reference values: $\tan\beta = 5$ (blue shapes)
and $\tan \beta = 20$ (red shapes). In the dark (blue and red) regions, we compare with current $\mu \to e \gamma$  bounds
while in the light (blue and red) regions we compare with the expected $\mu\to eee$ sensitivity in the near future.}
  \label{fig:S3current}
\end{figure}

The allowed parameter space for this model is given in Fig.~\ref{fig:S3current} for $\tan\beta=5$, blue (dark) areas, and $\tan\beta=20$, red (light) areas. Although we check all the low energy observables in Table~ \ref{tab:LFVbounds}, we once again find the most constraining processes to be $\mu \to e\gamma$ and $\mu \to e e e$. As can be seen in the figure, for low values of $\tan\beta$ this model seems to be slightly more constrained than $A_4$ whereas, for $\tan\beta=20$, the limits are practically the same. As in the $A_4$ case, the dominant contributions to these processes come from the LL sector and therefore an analogous description holds here: the LL leading terms, $\tan\beta$-enhanced, are those corresponding to contributions with an internal chirality flip and no cancellation among these terms occurs, since no relative sign from the hypercharge is present.
\\


\section{Conclusions} \label{conclusions}

In this work, building on the methods of \cite{Das:2016czs}, we continue to analyze the flavor structures in supersymmetric theories where the MSSM arises as a low energy effective theory from a flavor symmetry broken at higher scales. For a specific class of predictive models, if the scale of mediation of Supersymmetry breaking is above the flavor symmetry scale, the resulting flavor structures in the soft-breaking terms are not universal and can give rise to flavor changing effects at low energies.

We have applied these ideas to three representative discrete flavor symmetry models, $A_4$,$S_3$, and $\Delta(27)$, able to explain the neutrino and charged lepton structures. In these models, we have been able to obtain the full trilinear couplings and the soft mass matrices and we have applied the constraints from the non-observation of lepton flavor violating processes, like $\mu \to e \gamma$ and $\mu \to e e e$. We saw that different models may be distinguished through the different predicted structures in the trilinear terms or soft mass matrices. We have shown that, at present, these constraints are already competitive with direct LHC searches. Future bounds on these observables may discover SUSY with masses far beyond the reach of the LHC high-luminosity upgrade.

In conclusion, flavor symmetries in a supersymmetric context give rise generically to non-universal soft-breaking terms. This non-universality and the resulting flavor-changing effects must be always taken into account when restricting the allowed parameter space in these models. Moreover, the power of flavor changing observables to signal the presence of supersymmetry at higher scales has been explicitly demonstrated in these calculable models. We hope to continue to extend these results to unified models with symmetries that describe both the quark and lepton sectors in a future work. 
\appendix
\section{$A_4$ group}
\label{app:A4group}

The set of even permutations on four objects form a group, labeled $A_4$. This group can be generated by two elements $S$ and $T$ obeying the following relations 
\begin{equation}
S^2=(ST)^3=T^3=1.
\end{equation}

It has three independent one-dimensional representations $\mathbf{1}$,$\mathbf{1'}$,$\mathbf{1''}$ and one three-dimensional representation $\mathbf{3}$. The one-dimensional representations are given by:
\begin{equation}
\begin{split}
&\mathbf{1}\hspace{4mm}S=1 \hspace{4mm} T=1\\
&\mathbf{1'}\hspace{4mm}S=1 \hspace{4mm} T=e^{i4\pi/3}=\omega^2\\
&\mathbf{1''} \hspace{4mm}S=1 \hspace{4mm}  T=e^{i2\pi/3}=\omega.
\end{split}
\end{equation}
The three-dimensional representation, in a basis where the generator $T$ is diagonal, is given by:
\begin{equation}
T = \left(\begin{array}{ccc} 1 & 0 & 0\\ 0 & \omega^2& 0\\ 0 & 0 & \omega\\\end{array}\right)\hspace{5mm},\hspace{5mm}
S = \frac{1}{3}\left(\begin{array}{ccc} -1 & 2 & 2\\ 2 & -1 & 2\\ 2 & 2 & -1\\\end{array}\right).
\end{equation}
The multiplication rules between the various representations are:
\begin{equation}
\begin{split}
&\mathbf{1}\otimes\mathbf{1^{any}}=\mathbf{1^{any}}\hspace{4mm},\hspace{4mm}\mathbf{1'}\otimes\mathbf{1^{'}}=\mathbf{1^{''}}\hspace{4mm},\hspace{4mm}\mathbf{1'}\otimes\mathbf{1''}=\mathbf{1}\hspace{4mm},\hspace{4mm}\mathbf{1''}\otimes\mathbf{1''}=\mathbf{1'},
\end{split}
\end{equation}
then, taking ${\bf 3}_\alpha=(\alpha_1,\alpha_2,\alpha_3)$ and ${\bf 3}_\beta=(\beta_1,\beta_2,\beta_3)$ as two generic triplets, we can write also
\begin{equation}
\mathbf{1}\otimes\mathbf{3}_\alpha = \mathbf{3}_\alpha \sim \left(\begin{array}{c} \alpha_1 \\ \alpha_2 \\ \alpha_3\end{array}\right)\hspace{5mm},\hspace{5mm}
\mathbf{1'}\otimes\mathbf{3}_\alpha = \mathbf{3} \sim \left(\begin{array}{c} \alpha_3 \\ \alpha_1 \\ \alpha_2\end{array}\right)\hspace{5mm},\hspace{5mm}
\mathbf{1''}\otimes\mathbf{3}_\alpha = \mathbf{3} \sim \left(\begin{array}{c} \alpha_2 \\ \alpha_3 \\ \alpha_1\end{array}\right)
\end{equation}
\begin{equation}
\mathbf{3}_\alpha \otimes\mathbf{3}_\beta = \mathbf{1}+\mathbf{1'}+\mathbf{1''}+\mathbf{3}_S+\mathbf{3}_A \hspace{4mm}\mathrm{with}\hspace{4mm}  
\left\lbrace
\begin{array}{c} 
\mathbf{1}\sim \alpha_1\beta_1+\alpha_2\beta_3+\alpha_3\beta_2 \\ 
\mathbf{1'} \sim \alpha_3\beta_3+\alpha_1\beta_2+\alpha_2\beta_1 \\
\mathbf{1''} \sim \alpha_2\beta_2+\alpha_1\beta_3+\alpha_3\beta_1\\
\cr
\mathbf{3}_S \sim \frac{1}{3}\left(\begin{array}{c} 2\alpha_1\beta_1-\alpha_2\beta_3-\alpha_3\beta_2 \\2\alpha_3\beta_3-\alpha_1\beta_2-\alpha_2\beta_1 \\ 2\alpha_2\beta_2-\alpha_1\beta_3-\alpha_3\beta_1\end{array}\right)\\
\cr
\mathbf{3}_A \sim \frac{1}{2}\left(\begin{array}{c} \alpha_2\beta_3-\alpha_3\beta_2 \\\alpha_1\beta_2-\alpha_2\beta_1 \\ \alpha_1\beta_3-\alpha_3\beta_1\end{array}\right)\end{array}.
\right.
\end{equation}
It is useful to note that the operation of complex conjugation acts as
\begin{equation}
\begin{split}
&{\mathbf{1}}^*\sim\mathbf{1}\hspace{4mm},\hspace{4mm}(\mathbf{1'})^*\sim\mathbf{1''}\hspace{4mm},\hspace{4mm}(\mathbf{1''})^*\sim \mathbf{1'}\hspace{4mm},\hspace{4mm}\mathbf{3}^* \sim \left(\begin{array}{c} {\alpha_1}^* \\ {\alpha_3}^* \\ {\alpha_2}^*\end{array} \right),
\end{split}
\end{equation}
so, for example, the product rule $(\mathbf{1'} \otimes \mathbf{3})^*=\mathbf{1''} \otimes  \mathbf{3}^*$.The reason for this is that $T^*=U^T_{23}TU_{23}$ and $S^*=U^T_{23}SU_{23}=S$ where $U_{23}$ is the matrix that changes the 2nd and 3rd row and column.

\section{$S_3$ group}
\label{app:S3group}

The group $S_3$ is defined by the possible permutations among three objects. One of its presentations is that given by the generators $S$ and $T$ satisfying the following relations 
\begin{equation}
S^2=(ST)^2=T^3=1.
\end{equation}

The number of irreducible representations is three: two one-dimensional, $\mathbf{1}$ and $\mathbf{1'}$, and one two-dimensional, $\mathbf{2}$. The generators in the one-dimensional representations are given by:
\begin{equation}
\begin{split}
&\mathbf{1}\hspace{4mm}S=1 \hspace{4mm} T=1\\
&\mathbf{1'}\hspace{4mm}S=-1 \hspace{4mm} T=1\\
\end{split}
\end{equation}
while, in the two-dimensional representation for the T-diagonal basis, they can be written as:
\begin{equation}
T = \left(\begin{array}{cc} \omega & 0 \\ 0 & \omega^2 \\ \end{array}\right)\hspace{5mm},\hspace{5mm}
S = \left(\begin{array}{cc} 0 & 1 \\ 1 & 0 \\ \end{array}\right).
\end{equation}
The tensor products between singlets and pseudosinglets are:
\begin{equation}
\begin{split}
&\mathbf{1}\otimes\mathbf{1^{any}}=\mathbf{1^{any}}\hspace{4mm},\hspace{4mm}\mathbf{1'}\otimes\mathbf{1^{'}}=\mathbf{1}
\end{split}
\end{equation}
Considering two doublets, ${\bf 2}_\alpha=(\alpha_1,\alpha_2)$ and ${\bf 2}_\beta=(\beta_1,\beta_2)$, we can also write
\begin{equation}
\mathbf{1}\otimes\mathbf{2}_\alpha = \mathbf{2}_\alpha \sim \left(\begin{array}{c} \alpha_1 \\ \alpha_2\end{array}\right)\hspace{5mm},\hspace{5mm}
\mathbf{1'}\otimes\mathbf{2}_\alpha = \mathbf{2} \sim \left(\begin{array}{c} -\alpha_1 \\ \alpha_2 \end{array}\right)\hspace{5mm}
\end{equation}
\begin{equation}
\mathbf{2}_\alpha\otimes\mathbf{2}_\beta = \mathbf{1}+\mathbf{1'}+\mathbf{2} \hspace{4mm}\mathrm{with}\hspace{4mm}  
\left\lbrace
\begin{array}{c} 
\mathbf{1}\sim \alpha_1\beta_2+\alpha_2\beta_1 \\ 
\cr
\mathbf{1'} \sim \alpha_1\beta_2-\alpha_2\beta_1 \\
\cr
\mathbf{2} \sim \left(\begin{array}{c} \alpha_2\beta_2 \\ \alpha_1\beta_1 \end{array}\right)\\
\end{array}\right.
\end{equation}

The operation of complex conjugation leaves the singlets unchanged but acts over the doublet as follows
\begin{equation}
\begin{split}
\mathbf{2}^* \sim \left(\begin{array}{c} {\alpha_2}^* \\ {\alpha_1}^*\end{array} \right),
\end{split}
\end{equation}
so that ${\bf 2}^*$ transforms now as an anti-doublet with the matrices $(S^*,\, T^*)$.

\acknowledgments
This work has been partially supported under MINECO Grant FPA2014-57816-P and by the ``Centro de Excelencia Severo Ochoa'' Program under grant SEV-2014-0398. A. M. acknowledges support from ``La Caixa-Severo Ochoa'' scholarship. All Feynman diagrams have been drawn using Jaxodraw\cite{Binosi:2003yf,Binosi:2008ig}.

\bibliographystyle{JHEP}
\bibliography{references.bib}

\providecommand{\href}[2]{#2}\begingroup\raggedright\begin{thebibliography}{10}

\bibitem{Glashow:1970gm}
S.~L. Glashow, J.~Iliopoulos and L.~Maiani, \emph{{Weak Interactions with
  Lepton-Hadron Symmetry}},
  \href{https://doi.org/10.1103/PhysRevD.2.1285}{\emph{Phys. Rev.} {\bfseries
  D2} (1970) 1285--1292}.

\bibitem{Babu:2009fd}
K.~S. Babu, \emph{{TASI Lectures on Flavor Physics}},  in \emph{{Proceedings of
  Theoretical Advanced Study Institute in Elementary Particle Physics on The
  dawn of the LHC era (TASI 2008)}}, pp.~49--123, 2010,
  \href{https://arxiv.org/abs/0910.2948}{{\ttfamily 0910.2948}},
  \href{https://doi.org/10.1142/9789812838360_0002}{DOI}.

\bibitem{Altarelli:2010gt}
G.~Altarelli and F.~Feruglio, \emph{{Discrete Flavor Symmetries and Models of
  Neutrino Mixing}},
  \href{https://doi.org/10.1103/RevModPhys.82.2701}{\emph{Rev. Mod. Phys.}
  {\bfseries 82} (2010) 2701--2729},
  [\href{https://arxiv.org/abs/1002.0211}{{\ttfamily 1002.0211}}].

\bibitem{Das:2016czs}
D.~Das, M.~L. L\'opez-Ibáñez, M.~J. P\'erez and O.~Vives, \emph{{Effective
  theories of flavor and the nonuniversal MSSM}},
  \href{https://doi.org/10.1103/PhysRevD.95.035001}{\emph{Phys. Rev.}
  {\bfseries D95} (2017) 035001},
  [\href{https://arxiv.org/abs/1607.06827}{{\ttfamily 1607.06827}}].

\bibitem{Calibbi:2012yj}
L.~Calibbi, Z.~Lalak, S.~Pokorski and R.~Ziegler, \emph{{The Messenger Sector
  of SUSY Flavour Models and Radiative Breaking of Flavour Universality}},
  \href{https://doi.org/10.1007/JHEP06(2012)018}{\emph{JHEP} {\bfseries 06}
  (2012) 018}, [\href{https://arxiv.org/abs/1203.1489}{{\ttfamily 1203.1489}}].

\bibitem{Antusch:2011sq}
S.~Antusch, L.~Calibbi, V.~Maurer and M.~Spinrath, \emph{{From Flavour to SUSY
  Flavour Models}},
  \href{https://doi.org/10.1016/j.nuclphysb.2011.06.022}{\emph{Nucl. Phys.}
  {\bfseries B852} (2011) 108--148},
  [\href{https://arxiv.org/abs/1104.3040}{{\ttfamily 1104.3040}}].

\bibitem{King:2004tx}
S.~F. King, I.~N.~R. Peddie, G.~G. Ross, L.~Velasco-Sevilla and O.~Vives,
  \emph{{Kahler corrections and softly broken family symmetries}},
  \href{https://doi.org/10.1088/1126-6708/2005/07/049}{\emph{JHEP} {\bfseries
  07} (2005) 049}, [\href{https://arxiv.org/abs/hep-ph/0407012}{{\ttfamily
  hep-ph/0407012}}].

\bibitem{Olive:2016xmw}
{\scshape Particle Data Group} collaboration, C.~Patrignani et~al.,
  \emph{{Review of Particle Physics}},
  \href{https://doi.org/10.1088/1674-1137/40/10/100001}{\emph{Chin. Phys.}
  {\bfseries C40} (2016) 100001}.

\bibitem{Baldini:2013ke}
A.~M. Baldini et~al., \emph{{MEG Upgrade Proposal}},
  \href{https://arxiv.org/abs/1301.7225}{{\ttfamily 1301.7225}}.

\bibitem{Aushev:2010bq}
T.~Aushev et~al., \emph{{Physics at Super B Factory}},
  \href{https://arxiv.org/abs/1002.5012}{{\ttfamily 1002.5012}}.

\bibitem{Blondel:2013ia}
A.~Blondel et~al., \emph{{Research Proposal for an Experiment to Search for the
  Decay $\mu \to eee$}},  \href{https://arxiv.org/abs/1301.6113}{{\ttfamily
  1301.6113}}.

\bibitem{Casas:1996de}
J.~A. Casas and S.~Dimopoulos, \emph{{Stability bounds on flavor violating
  trilinear soft terms in the MSSM}},
  \href{https://doi.org/10.1016/0370-2693(96)01000-3}{\emph{Phys. Lett.}
  {\bfseries B387} (1996) 107--112},
  [\href{https://arxiv.org/abs/hep-ph/9606237}{{\ttfamily hep-ph/9606237}}].

\bibitem{deMedeirosVarzielas:2006fc}
I.~de~Medeiros~Varzielas, S.~F. King and G.~G. Ross, \emph{{Neutrino
  tri-bi-maximal mixing from a non-Abelian discrete family symmetry}},
  \href{https://doi.org/10.1016/j.physletb.2007.03.009}{\emph{Phys. Lett.}
  {\bfseries B648} (2007) 201--206},
  [\href{https://arxiv.org/abs/hep-ph/0607045}{{\ttfamily hep-ph/0607045}}].

\bibitem{deMedeirosVarzielas:2005ax}
I.~de~Medeiros~Varzielas and G.~G. Ross, \emph{{SU(3) family symmetry and
  neutrino bi-tri-maximal mixing}},
  \href{https://doi.org/10.1016/j.nuclphysb.2005.10.039}{\emph{Nucl. Phys.}
  {\bfseries B733} (2006) 31--47},
  [\href{https://arxiv.org/abs/hep-ph/0507176}{{\ttfamily hep-ph/0507176}}].

\bibitem{deMedeirosVarzielas:2017sdv}
I.~de~Medeiros~Varzielas, G.~G. Ross and J.~Talbert, \emph{{A Unified Model of
  Quarks and Leptons with a Universal Texture Zero}},
  \href{https://arxiv.org/abs/1710.01741}{{\ttfamily 1710.01741}}.

\bibitem{Patrignani:2016xqp}
{\scshape Particle Data Group} collaboration, C.~Patrignani et~al.,
  \emph{{Review of Particle Physics}},
  \href{https://doi.org/10.1088/1674-1137/40/10/100001}{\emph{Chin. Phys.}
  {\bfseries C40} (2016) 100001}.

\bibitem{Antusch:2013jca}
S.~Antusch and V.~Maurer, \emph{{Running quark and lepton parameters at various
  scales}}, \href{https://doi.org/10.1007/JHEP11(2013)115}{\emph{JHEP}
  {\bfseries 11} (2013) 115},
  [\href{https://arxiv.org/abs/1306.6879}{{\ttfamily 1306.6879}}].

\bibitem{Porod:2011nf}
W.~Porod and F.~Staub, \emph{{SPheno 3.1: Extensions including flavour,
  CP-phases and models beyond the MSSM}},
  \href{https://doi.org/10.1016/j.cpc.2012.05.021}{\emph{Comput. Phys. Commun.}
  {\bfseries 183} (2012) 2458--2469},
  [\href{https://arxiv.org/abs/1104.1573}{{\ttfamily 1104.1573}}].

\bibitem{Porod:2003um}
W.~Porod, \emph{{SPheno, a program for calculating supersymmetric spectra, SUSY
  particle decays and SUSY particle production at e+ e- colliders}},
  \href{https://doi.org/10.1016/S0010-4655(03)00222-4}{\emph{Comput. Phys.
  Commun.} {\bfseries 153} (2003) 275--315},
  [\href{https://arxiv.org/abs/hep-ph/0301101}{{\ttfamily hep-ph/0301101}}].

\bibitem{Hall:1985dx}
L.~J. Hall, V.~A. Kostelecky and S.~Raby, \emph{{New Flavor Violations in
  Supergravity Models}},
  \href{https://doi.org/10.1016/0550-3213(86)90397-4}{\emph{Nucl. Phys.}
  {\bfseries B267} (1986) 415--432}.

\bibitem{Gabbiani:1988rb}
F.~Gabbiani and A.~Masiero, \emph{{FCNC in Generalized Supersymmetric
  Theories}}, \href{https://doi.org/10.1016/0550-3213(89)90492-6}{\emph{Nucl.
  Phys.} {\bfseries B322} (1989) 235--254}.

\bibitem{Gabbiani:1996hi}
F.~Gabbiani, E.~Gabrielli, A.~Masiero and L.~Silvestrini, \emph{A complete
  analysis of fcnc and cp constraints in general susy extensions of the
  standard model}, {\emph{Nucl. Phys.} {\bfseries B477} (1996) 321--352},
  [\href{https://arxiv.org/abs/hep-ph/9604387}{{\ttfamily hep-ph/9604387}}].

\bibitem{Paradisi:2005fk}
P.~Paradisi, \emph{{Constraints on SUSY lepton flavor violation by rare
  processes}}, \href{https://doi.org/10.1088/1126-6708/2005/10/006}{\emph{JHEP}
  {\bfseries 10} (2005) 006},
  [\href{https://arxiv.org/abs/hep-ph/0505046}{{\ttfamily hep-ph/0505046}}].

\bibitem{Ciuchini:2007ha}
M.~Ciuchini, A.~Masiero, P.~Paradisi, L.~Silvestrini, S.~K. Vempati and
  O.~Vives, \emph{{Soft SUSY breaking grand unification: Leptons versus quarks
  on the flavor playground}},
  \href{https://doi.org/10.1016/j.nuclphysb.2007.05.032}{\emph{Nucl. Phys.}
  {\bfseries B783} (2007) 112--142},
  [\href{https://arxiv.org/abs/hep-ph/0702144}{{\ttfamily hep-ph/0702144}}].

\bibitem{Calibbi:2008qt}
L.~Calibbi, J.~Jones-Perez and O.~Vives, \emph{{Electric dipole moments from
  flavoured CP violation in SUSY}},
  \href{https://doi.org/10.1103/PhysRevD.78.075007}{\emph{Phys. Rev.}
  {\bfseries D78} (2008) 075007},
  [\href{https://arxiv.org/abs/0804.4620}{{\ttfamily 0804.4620}}].

\bibitem{Calibbi:2009ja}
L.~Calibbi, J.~Jones-Perez, A.~Masiero, J.-h. Park, W.~Porod and O.~Vives,
  \emph{{FCNC and CP Violation Observables in a SU(3)-flavoured MSSM}},
  \href{https://doi.org/10.1016/j.nuclphysb.2009.12.029}{\emph{Nucl. Phys.}
  {\bfseries B831} (2010) 26--71},
  [\href{https://arxiv.org/abs/0907.4069}{{\ttfamily 0907.4069}}].

\bibitem{Altarelli:2005yx}
G.~Altarelli and F.~Feruglio, \emph{{Tri-bimaximal neutrino mixing, A(4) and
  the modular symmetry}},
  \href{https://doi.org/10.1016/j.nuclphysb.2006.02.015}{\emph{Nucl. Phys.}
  {\bfseries B741} (2006) 215--235},
  [\href{https://arxiv.org/abs/hep-ph/0512103}{{\ttfamily hep-ph/0512103}}].

\bibitem{Altarelli:2009kr}
G.~Altarelli and D.~Meloni, \emph{{A Simplest A4 Model for Tri-Bimaximal
  Neutrino Mixing}},
  \href{https://doi.org/10.1088/0954-3899/36/8/085005}{\emph{J. Phys.}
  {\bfseries G36} (2009) 085005},
  [\href{https://arxiv.org/abs/0905.0620}{{\ttfamily 0905.0620}}].

\bibitem{Altarelli:2005yp}
G.~Altarelli and F.~Feruglio, \emph{{Tri-bimaximal neutrino mixing from
  discrete symmetry in extra dimensions}},
  \href{https://doi.org/10.1016/j.nuclphysb.2005.05.005}{\emph{Nucl. Phys.}
  {\bfseries B720} (2005) 64--88},
  [\href{https://arxiv.org/abs/hep-ph/0504165}{{\ttfamily hep-ph/0504165}}].

\bibitem{Ma:2001dn}
E.~Ma and G.~Rajasekaran, \emph{{Softly broken A(4) symmetry for nearly
  degenerate neutrino masses}},
  \href{https://doi.org/10.1103/PhysRevD.64.113012}{\emph{Phys. Rev.}
  {\bfseries D64} (2001) 113012},
  [\href{https://arxiv.org/abs/hep-ph/0106291}{{\ttfamily hep-ph/0106291}}].

\bibitem{Babu:2002dz}
K.~S. Babu, E.~Ma and J.~W.~F. Valle, \emph{{Underlying A(4) symmetry for the
  neutrino mass matrix and the quark mixing matrix}},
  \href{https://doi.org/10.1016/S0370-2693(02)03153-2}{\emph{Phys. Lett.}
  {\bfseries B552} (2003) 207--213},
  [\href{https://arxiv.org/abs/hep-ph/0206292}{{\ttfamily hep-ph/0206292}}].

\bibitem{Ma:2004zv}
E.~Ma, \emph{{A(4) symmetry and neutrinos with very different masses}},
  \href{https://doi.org/10.1103/PhysRevD.70.031901}{\emph{Phys. Rev.}
  {\bfseries D70} (2004) 031901},
  [\href{https://arxiv.org/abs/hep-ph/0404199}{{\ttfamily hep-ph/0404199}}].

\bibitem{He:2006dk}
X.-G. He, Y.-Y. Keum and R.~R. Volkas, \emph{{A(4) flavor symmetry breaking
  scheme for understanding quark and neutrino mixing angles}},
  \href{https://doi.org/10.1088/1126-6708/2006/04/039}{\emph{JHEP} {\bfseries
  04} (2006) 039}, [\href{https://arxiv.org/abs/hep-ph/0601001}{{\ttfamily
  hep-ph/0601001}}].

\bibitem{Babu:2005se}
K.~S. Babu and X.-G. He, \emph{{Model of geometric neutrino mixing}},
  \href{https://arxiv.org/abs/hep-ph/0507217}{{\ttfamily hep-ph/0507217}}.

\bibitem{Zee:2005ut}
A.~Zee, \emph{{Obtaining the neutrino mixing matrix with the tetrahedral
  group}}, \href{https://doi.org/10.1016/j.physletb.2005.09.068}{\emph{Phys.
  Lett.} {\bfseries B630} (2005) 58--67},
  [\href{https://arxiv.org/abs/hep-ph/0508278}{{\ttfamily hep-ph/0508278}}].

\bibitem{Ma:2005qf}
E.~Ma, \emph{{Tribimaximal neutrino mixing from a supersymmetric model with A4
  family symmetry}},
  \href{https://doi.org/10.1103/PhysRevD.73.057304}{\emph{Phys. Rev.}
  {\bfseries D73} (2006) 057304},
  [\href{https://arxiv.org/abs/hep-ph/0511133}{{\ttfamily hep-ph/0511133}}].

\bibitem{King:2006np}
S.~F. King and M.~Malinsky, \emph{{A(4) family symmetry and quark-lepton
  unification}},
  \href{https://doi.org/10.1016/j.physletb.2006.12.006}{\emph{Phys. Lett.}
  {\bfseries B645} (2007) 351--357},
  [\href{https://arxiv.org/abs/hep-ph/0610250}{{\ttfamily hep-ph/0610250}}].

\bibitem{Altarelli:2006kg}
G.~Altarelli, F.~Feruglio and Y.~Lin, \emph{{Tri-bimaximal neutrino mixing from
  orbifolding}},
  \href{https://doi.org/10.1016/j.nuclphysb.2007.03.042}{\emph{Nucl. Phys.}
  {\bfseries B775} (2007) 31--44},
  [\href{https://arxiv.org/abs/hep-ph/0610165}{{\ttfamily hep-ph/0610165}}].

\bibitem{Hirsch:2007kh}
M.~Hirsch, A.~S. Joshipura, S.~Kaneko and J.~W.~F. Valle, \emph{{Predictive
  flavour symmetries of the neutrino mass matrix}},
  \href{https://doi.org/10.1103/PhysRevLett.99.151802}{\emph{Phys. Rev. Lett.}
  {\bfseries 99} (2007) 151802},
  [\href{https://arxiv.org/abs/hep-ph/0703046}{{\ttfamily hep-ph/0703046}}].

\bibitem{Bazzocchi:2007na}
F.~Bazzocchi, S.~Kaneko and S.~Morisi, \emph{{A SUSY A(4) model for fermion
  masses and mixings}},
  \href{https://doi.org/10.1088/1126-6708/2008/03/063}{\emph{JHEP} {\bfseries
  03} (2008) 063}, [\href{https://arxiv.org/abs/0707.3032}{{\ttfamily
  0707.3032}}].

\bibitem{Honda:2008rs}
M.~Honda and M.~Tanimoto, \emph{{Deviation from tri-bimaximal neutrino mixing
  in A(4) flavor symmetry}},
  \href{https://doi.org/10.1143/PTP.119.583}{\emph{Prog. Theor. Phys.}
  {\bfseries 119} (2008) 583--598},
  [\href{https://arxiv.org/abs/0801.0181}{{\ttfamily 0801.0181}}].

\bibitem{Bazzocchi:2008rz}
F.~Bazzocchi, S.~Morisi, M.~Picariello and E.~Torrente-Lujan, \emph{{Embedding
  A(4) into SU(3) x U(1) flavor symmetry: Large neutrino mixing and fermion
  mass hierarchy in SO(10) GUT}},
  \href{https://doi.org/10.1088/0954-3899/36/1/015002}{\emph{J. Phys.}
  {\bfseries G36} (2009) 015002},
  [\href{https://arxiv.org/abs/0802.1693}{{\ttfamily 0802.1693}}].

\bibitem{Hirsch:2008rp}
M.~Hirsch, S.~Morisi and J.~W.~F. Valle, \emph{{Tri-bimaximal neutrino mixing
  and neutrinoless double beta decay}},
  \href{https://doi.org/10.1103/PhysRevD.78.093007}{\emph{Phys. Rev.}
  {\bfseries D78} (2008) 093007},
  [\href{https://arxiv.org/abs/0804.1521}{{\ttfamily 0804.1521}}].

\bibitem{Lin:2008aj}
Y.~Lin, \emph{{A Predictive A(4) model, Charged Lepton Hierarchy and
  Tri-bimaximal Sum Rule}},
  \href{https://doi.org/10.1016/j.nuclphysb.2008.12.025}{\emph{Nucl. Phys.}
  {\bfseries B813} (2009) 91--105},
  [\href{https://arxiv.org/abs/0804.2867}{{\ttfamily 0804.2867}}].

\bibitem{Csaki:2008qq}
C.~Csaki, C.~Delaunay, C.~Grojean and Y.~Grossman, \emph{{A Model of Lepton
  Masses from a Warped Extra Dimension}},
  \href{https://doi.org/10.1088/1126-6708/2008/10/055}{\emph{JHEP} {\bfseries
  10} (2008) 055}, [\href{https://arxiv.org/abs/0806.0356}{{\ttfamily
  0806.0356}}].

\bibitem{Feruglio:2008ht}
F.~Feruglio, C.~Hagedorn, Y.~Lin and L.~Merlo, \emph{{Lepton Flavour Violation
  in Models with A(4) Flavour Symmetry}},
  \href{https://doi.org/10.1016/j.nuclphysb.2008.10.002}{\emph{Nucl. Phys.}
  {\bfseries B809} (2009) 218--243},
  [\href{https://arxiv.org/abs/0807.3160}{{\ttfamily 0807.3160}}].

\bibitem{Morisi:2007ft}
S.~Morisi, M.~Picariello and E.~Torrente-Lujan, \emph{{Model for fermion masses
  and lepton mixing in SO(10) x A(4)}},
  \href{https://doi.org/10.1103/PhysRevD.75.075015}{\emph{Phys. Rev.}
  {\bfseries D75} (2007) 075015},
  [\href{https://arxiv.org/abs/hep-ph/0702034}{{\ttfamily hep-ph/0702034}}].

\bibitem{Hirsch:2005mc}
M.~Hirsch, A.~Villanova~del Moral, J.~W.~F. Valle and E.~Ma, \emph{{Predicting
  neutrinoless double beta decay}},
  \href{https://doi.org/10.1103/PhysRevD.72.091301,
  10.1103/PhysRevD.72.119904}{\emph{Phys. Rev.} {\bfseries D72} (2005) 091301},
  [\href{https://arxiv.org/abs/hep-ph/0507148}{{\ttfamily hep-ph/0507148}}].

\bibitem{Hirsch:2003dr}
M.~Hirsch, J.~C. Romao, S.~Skadhauge, J.~W.~F. Valle and A.~Villanova~del
  Moral, \emph{{Phenomenological tests of supersymmetric A(4) family symmetry
  model of neutrino mass}},
  \href{https://doi.org/10.1103/PhysRevD.69.093006}{\emph{Phys. Rev.}
  {\bfseries D69} (2004) 093006},
  [\href{https://arxiv.org/abs/hep-ph/0312265}{{\ttfamily hep-ph/0312265}}].

\bibitem{Chen:2009um}
M.-C. Chen and S.~F. King, \emph{{A4 See-Saw Models and Form Dominance}},
  \href{https://doi.org/10.1088/1126-6708/2009/06/072}{\emph{JHEP} {\bfseries
  06} (2009) 072}, [\href{https://arxiv.org/abs/0903.0125}{{\ttfamily
  0903.0125}}].

\bibitem{Hirsch:2003xx}
M.~Hirsch, J.~C. Romao, S.~Skadhauge, J.~W.~F. Valle and A.~Villanova~del
  Moral, \emph{{Degenerate neutrinos from a supersymmetric A(4) model}},
  \href{https://arxiv.org/abs/hep-ph/0312244}{{\ttfamily hep-ph/0312244}}.

\bibitem{Abe:2014bwa}
{\scshape Double Chooz} collaboration, Y.~Abe et~al., \emph{{Improved
  measurements of the neutrino mixing angle $\theta_{13}$ with the Double Chooz
  detector}}, \href{https://doi.org/10.1007/JHEP02(2015)074,
  10.1007/JHEP10(2014)086}{\emph{JHEP} {\bfseries 10} (2014) 086},
  [\href{https://arxiv.org/abs/1406.7763}{{\ttfamily 1406.7763}}].

\bibitem{An:2015rpe}
{\scshape Daya Bay} collaboration, F.~P. An et~al., \emph{{New Measurement of
  Antineutrino Oscillation with the Full Detector Configuration at Daya Bay}},
  \href{https://doi.org/10.1103/PhysRevLett.115.111802}{\emph{Phys. Rev. Lett.}
  {\bfseries 115} (2015) 111802},
  [\href{https://arxiv.org/abs/1505.03456}{{\ttfamily 1505.03456}}].

\bibitem{RENO:2015ksa}
{\scshape RENO} collaboration, J.~H. Choi et~al., \emph{{Observation of Energy
  and Baseline Dependent Reactor Antineutrino Disappearance in the RENO
  Experiment}},
  \href{https://doi.org/10.1103/PhysRevLett.116.211801}{\emph{Phys. Rev. Lett.}
  {\bfseries 116} (2016) 211801},
  [\href{https://arxiv.org/abs/1511.05849}{{\ttfamily 1511.05849}}].

\bibitem{Lin:2009bw}
Y.~Lin, \emph{{Tri-bimaximal Neutrino Mixing from A(4) and $\theta_{13}$ ~
  theta(C)}},
  \href{https://doi.org/10.1016/j.nuclphysb.2009.08.018}{\emph{Nucl. Phys.}
  {\bfseries B824} (2010) 95--110},
  [\href{https://arxiv.org/abs/0905.3534}{{\ttfamily 0905.3534}}].

\bibitem{Cooper:2012wf}
I.~K. Cooper, S.~F. King and C.~Luhn, \emph{{A4xSU(5) SUSY GUT of Flavour with
  Trimaximal Neutrino Mixing}},
  \href{https://doi.org/10.1007/JHEP06(2012)130}{\emph{JHEP} {\bfseries 06}
  (2012) 130}, [\href{https://arxiv.org/abs/1203.1324}{{\ttfamily 1203.1324}}].

\bibitem{Hernandez:2012ra}
D.~Hernandez and A.~{\relax Yu}. Smirnov, \emph{{Lepton mixing and discrete
  symmetries}}, \href{https://doi.org/10.1103/PhysRevD.86.053014}{\emph{Phys.
  Rev.} {\bfseries D86} (2012) 053014},
  [\href{https://arxiv.org/abs/1204.0445}{{\ttfamily 1204.0445}}].

\bibitem{King:2011zj}
S.~F. King and C.~Luhn, \emph{{Trimaximal neutrino mixing from vacuum alignment
  in A4 and S4 models}},
  \href{https://doi.org/10.1007/JHEP09(2011)042}{\emph{JHEP} {\bfseries 09}
  (2011) 042}, [\href{https://arxiv.org/abs/1107.5332}{{\ttfamily 1107.5332}}].

\bibitem{Zheng:2011uz}
Y.-j. Zheng and B.-Q. Ma, \emph{{Re-Evaluation of Neutrino Mixing Pattern
  According to Latest T2K result}},
  \href{https://doi.org/10.1140/epjp/i2012-12007-1}{\emph{Eur. Phys. J. Plus}
  {\bfseries 127} (2012) 7}, [\href{https://arxiv.org/abs/1106.4040}{{\ttfamily
  1106.4040}}].

\bibitem{Ma:2011yi}
E.~Ma and D.~Wegman, \emph{{Nonzero theta(13) for neutrino mixing in the
  context of A(4) symmetry}},
  \href{https://doi.org/10.1103/PhysRevLett.107.061803}{\emph{Phys. Rev. Lett.}
  {\bfseries 107} (2011) 061803},
  [\href{https://arxiv.org/abs/1106.4269}{{\ttfamily 1106.4269}}].

\bibitem{Rodejohann:2011uz}
W.~Rodejohann, H.~Zhang and S.~Zhou, \emph{{Systematic search for successful
  lepton mixing patterns with nonzero $\theta_{13}$}},
  \href{https://doi.org/10.1016/j.nuclphysb.2011.10.017}{\emph{Nucl. Phys.}
  {\bfseries B855} (2012) 592--607},
  [\href{https://arxiv.org/abs/1107.3970}{{\ttfamily 1107.3970}}].

\bibitem{Ahn:2011if}
Y.~H. Ahn, H.-Y. Cheng and S.~Oh, \emph{{An extension of tribimaximal lepton
  mixing}}, \href{https://doi.org/10.1103/PhysRevD.84.113007}{\emph{Phys. Rev.}
  {\bfseries D84} (2011) 113007},
  [\href{https://arxiv.org/abs/1107.4549}{{\ttfamily 1107.4549}}].

\bibitem{Kumar:2011vf}
S.~Kumar, \emph{{Implications of a class of neutrino mass matrices with texture
  zeros for non-zero $\theta_{13}$}},
  \href{https://doi.org/10.1103/PhysRevD.84.077301}{\emph{Phys. Rev.}
  {\bfseries D84} (2011) 077301},
  [\href{https://arxiv.org/abs/1108.2137}{{\ttfamily 1108.2137}}].

\bibitem{Gupta:2011ct}
S.~Gupta, A.~S. Joshipura and K.~M. Patel, \emph{{Minimal extension of
  tri-bimaximal mixing and generalized $Z_2$ X $Z_2$ symmetries}},
  \href{https://doi.org/10.1103/PhysRevD.85.031903}{\emph{Phys. Rev.}
  {\bfseries D85} (2012) 031903},
  [\href{https://arxiv.org/abs/1112.6113}{{\ttfamily 1112.6113}}].

\bibitem{Branco:2012vs}
G.~C. Branco, R.~Gonzalez~Felipe, F.~R. Joaquim and H.~Serodio,
  \emph{{Spontaneous leptonic CP violation and nonzero $\theta_{13}$}},
  \href{https://doi.org/10.1103/PhysRevD.86.076008}{\emph{Phys. Rev.}
  {\bfseries D86} (2012) 076008},
  [\href{https://arxiv.org/abs/1203.2646}{{\ttfamily 1203.2646}}].

\bibitem{Ahn:2012tv}
Y.~H. Ahn and S.~K. Kang, \emph{{Non-zero $\theta_{13}$ and CP violation in a
  model with $A_4$ flavor symmetry}},
  \href{https://doi.org/10.1103/PhysRevD.86.093003}{\emph{Phys. Rev.}
  {\bfseries D86} (2012) 093003},
  [\href{https://arxiv.org/abs/1203.4185}{{\ttfamily 1203.4185}}].

\bibitem{Altarelli:2008bg}
G.~Altarelli, F.~Feruglio and C.~Hagedorn, \emph{{A SUSY SU(5) Grand Unified
  Model of Tri-Bimaximal Mixing from A$_4$}},
  \href{https://doi.org/10.1088/1126-6708/2008/03/052}{\emph{JHEP} {\bfseries
  03} (2008) 052}, [\href{https://arxiv.org/abs/0802.0090}{{\ttfamily
  0802.0090}}].

\bibitem{Haba:2005ds}
N.~Haba and K.~Yoshioka, \emph{{Discrete flavor symmetry, dynamical mass
  textures, and grand unification}},
  \href{https://doi.org/10.1016/j.nuclphysb.2006.01.027}{\emph{Nucl. Phys.}
  {\bfseries B739} (2006) 254--284},
  [\href{https://arxiv.org/abs/hep-ph/0511108}{{\ttfamily hep-ph/0511108}}].

\bibitem{Morisi:2005fy}
S.~Morisi and M.~Picariello, \emph{{The Flavor physics in unified gauge theory
  from an S(3) x P discrete symmetry}},
  \href{https://doi.org/10.1007/s10773-006-9126-z}{\emph{Int. J. Theor. Phys.}
  {\bfseries 45} (2006) 1267--1277},
  [\href{https://arxiv.org/abs/hep-ph/0505113}{{\ttfamily hep-ph/0505113}}].

\bibitem{Chen:2004rr}
S.-L. Chen, M.~Frigerio and E.~Ma, \emph{{Large neutrino mixing and normal mass
  hierarchy: A Discrete understanding}},
  \href{https://doi.org/10.1103/PhysRevD.70.079905,
  10.1103/PhysRevD.70.073008}{\emph{Phys. Rev.} {\bfseries D70} (2004) 073008},
  [\href{https://arxiv.org/abs/hep-ph/0404084}{{\ttfamily hep-ph/0404084}}].

\bibitem{Xing:2010iu}
Z.-z. Xing, D.~Yang and S.~Zhou, \emph{{Broken $S_3$ Flavor Symmetry of Leptons
  and Quarks: Mass Spectra and Flavor Mixing Patterns}},
  \href{https://doi.org/10.1016/j.physletb.2010.05.045}{\emph{Phys. Lett.}
  {\bfseries B690} (2010) 304--310},
  [\href{https://arxiv.org/abs/1004.4234}{{\ttfamily 1004.4234}}].

\bibitem{Grimus:2005mu}
W.~Grimus and L.~Lavoura, \emph{{S(3) x Z(2) model for neutrino mass
  matrices}}, \href{https://doi.org/10.1088/1126-6708/2005/08/013}{\emph{JHEP}
  {\bfseries 08} (2005) 013},
  [\href{https://arxiv.org/abs/hep-ph/0504153}{{\ttfamily hep-ph/0504153}}].

\bibitem{Mohapatra:2006pu}
R.~N. Mohapatra, S.~Nasri and H.-B. Yu, \emph{{S(3) symmetry and tri-bimaximal
  mixing}}, \href{https://doi.org/10.1016/j.physletb.2006.06.032}{\emph{Phys.
  Lett.} {\bfseries B639} (2006) 318--321},
  [\href{https://arxiv.org/abs/hep-ph/0605020}{{\ttfamily hep-ph/0605020}}].

\bibitem{Feruglio:2007hi}
F.~Feruglio and Y.~Lin, \emph{{Fermion Mass Hierarchies and Flavour Mixing from
  a Minimal Discrete Symmetry}},
  \href{https://doi.org/10.1016/j.nuclphysb.2008.02.008}{\emph{Nucl. Phys.}
  {\bfseries B800} (2008) 77--93},
  [\href{https://arxiv.org/abs/0712.1528}{{\ttfamily 0712.1528}}].

\bibitem{Ma:1991eg}
E.~Ma, \emph{{S(3) Z(3) model of lepton mass matrices}},
  \href{https://doi.org/10.1103/PhysRevD.44.587}{\emph{Phys. Rev.} {\bfseries
  D44} (1991) 587--589}.

\bibitem{Meloni:2012ci}
D.~Meloni, \emph{{$S_3$ as a flavour symmetry for quarks and leptons after the
  Daya Bay result on $\theta_{13}$}},
  \href{https://doi.org/10.1007/JHEP05(2012)124}{\emph{JHEP} {\bfseries 05}
  (2012) 124}, [\href{https://arxiv.org/abs/1203.3126}{{\ttfamily 1203.3126}}].

\bibitem{Binosi:2003yf}
D.~Binosi and L.~Theussl, \emph{{JaxoDraw: A Graphical user interface for
  drawing Feynman diagrams}},
  \href{https://doi.org/10.1016/j.cpc.2004.05.001}{\emph{Comput. Phys. Commun.}
  {\bfseries 161} (2004) 76--86},
  [\href{https://arxiv.org/abs/hep-ph/0309015}{{\ttfamily hep-ph/0309015}}].

\bibitem{Binosi:2008ig}
D.~Binosi, J.~Collins, C.~Kaufhold and L.~Theussl, \emph{{JaxoDraw: A Graphical
  user interface for drawing Feynman diagrams. Version 2.0 release notes}},
  \href{https://doi.org/10.1016/j.cpc.2009.02.020}{\emph{Comput. Phys. Commun.}
  {\bfseries 180} (2009) 1709--1715},
  [\href{https://arxiv.org/abs/0811.4113}{{\ttfamily 0811.4113}}].

\end{thebibliography}\endgroup
\end{document}